\documentclass[journal=jctcce,manuscript=article]{achemso}
\usepackage{chemformula} % Formula subscripts using \ch{}
\usepackage[T1]{fontenc} % Use modern font encodings
\usepackage{hyperref}
\usepackage{caption,subcaption}
\usepackage{physics,siunitx}
\usepackage{amsmath,amssymb,amsthm,bm}
\usepackage{tabularx}

\author{Carlos Floyd}
\affiliation{Department of Chemistry \& Biochemistry,
Institute for Physical Science and Technology,
University of Maryland,
College Park, MD 20742, USA}
\author{Haoran Ni}
\affiliation{Department of Chemistry \& Biochemistry,
Institute for Physical Science and Technology,
University of Maryland,
College Park, MD 20742, USA}
\author{Ravinda S. Gunaratne}
\affiliation{Mathematical Institute, University of Oxford, Radcliffe Observatory Quarter, Woodstock Road, Oxford, OX2 6GG, United Kingdom}
\author{Radek Erban}
\email{erban@maths.ox.ac.uk}
\affiliation{Mathematical Institute, University of Oxford, Radcliffe Observatory Quarter, Woodstock Road, Oxford, OX2 6GG, United Kingdom}
\author{Garegin A. Papoian}
\email{gpapoian@umd.edu}
\affiliation{Department of Chemistry \& Biochemistry,
Institute for Physical Science and Technology,
University of Maryland,
College Park, MD 20742, USA} 

\title{On Stretching, Bending, Shearing and Twisting of Actin Filaments I: \break Variational Models}

\begin{document}
\clearpage
\singlespacing

\begin{abstract}
\noindent
 Mechanochemical simulations of actomyosin networks are traditionally based on one-dimensional models of actin filaments having zero width. Here, and in the follow up paper, approaches are presented for more efficient modelling which incorporates stretching, bending, shearing and twisting of actin filaments. Our modelling of a semi-flexible filament with a small but finite width is based on the Cosserat theory of elastic rods, which allows for six degrees of freedom at every point on the filament's backbone. In the variational models presented in this paper, a small and discrete set of parameters is used to describe a smooth filament shape having all degrees of freedom allowed in the Cosserat theory. Two main approaches are introduced: one where polynomial spline functions describe the filament's configuration, and one in which geodesic curves in the space of the configurational degrees of freedom are used. We find that in the latter representation the strain energy function can be calculated without resorting to a small-angle expansion, so it can describe arbitrarily large filament deformations without systematic error. These approaches are validated by a dynamical model of a Cosserat filament, which can be further extended by using multi-resolution methods to allow more detailed monomer-based resolution in certain parts of the actin filament, as introduced in the follow up paper. The presented framework is illustrated by showing how torsional compliance in a finite-width filament can induce broken chiral symmetry in the structure of a cross-linked bundle. 
\end{abstract}

\clearpage
\section{Introduction}\label{Introduction}

Simulations of the actin-based cytoskeleton allow for deep insights into its dynamics and mechanical properties.  Composed primarily of cross-linked actin filaments and molecular motors, this structural protein network exhibits fascinating behaviors on a range of spatial scales~\cite{fletcher2010cell,howard2001mechanics}.  From the \r{a}ngstr\"om scale, at which individual actin monomers and molecular motors hydrolyze chemical fuel to drive conformational changes, up to the millimeter scale, at which collectives of cells exert self-organized mutual forces on one another, the nonequilibrium dynamics and mechanics of cytoskeletal networks enable much of the cellular functionality necessary for life~\cite{mccullagh2014unraveling,belmont1999change,mani2021compressive,mizuno2007nonequilibrium,cordes2020prestress,gardel2004elastic,ajeti2019wound,floyd2019quantifying}.  Associated with this wide range of spatial scales is a variety of computational techniques that are used for modeling cytoskeletal networks~\cite{yamaoka2012multiscale}.  Each technique accounts to some level of approximation for the mechanical and geometrical properties of actin filaments.  These filaments can be classified as semi-flexible polymers (whose typical contour lengths are comparable to their persistence length) with very large aspect ratios (such that the contour length is much greater than the filament's radius)~\cite{broedersz2014modeling,rubinstein2003polymer}. In this paper, we will focus on network-level computational models of cytoskeletal networks, which typically assign $\sim 10 - 100$ monomers to a single discrete computational element.  Software packages such as AFiNeS~\cite{freedman2017versatile}, CytoSim~\cite{nedelec2007collective}, the model of Kim and coworkers~\cite{kim2009computational}, and MEDYAN~\cite{popov2016medyan} can access time scales of thousands of seconds and length scales of tens of micrometers, allowing exploration of fascinating emergent phenomena of cytoskeletal systems which comprise many interacting filaments.  In the coarse-grained mechanical models used in these platforms, an actin filament is represented as a one-dimensional piecewise-linear chain of elastic segments with stretching and bending energy penalties.  An effective radius can be assigned to the filament so that it experiences excluded volume interactions with its neighbors to prevent overlap, but the elastic strain energy functions used in these models neglect the filament width~\cite{floyd2021segmental}.

There is strong reason to expect that the finite width of an actin filament, neglected in current network-level models of cytoskeletal systems, plays an important role in cytoskeletal dynamics. Experiments \textit{in vivo} and \textit{in vitro} have illustrated the emergence of remarkable rotating dynamical phases of cytoskeletal systems, in which vortex structures spontaneously emerge as a broken chiral symmetry of the system~\cite{tee2015cellular, schaller2010polar,fritzsche2017self}. These collective rotating phases likely involve torques exerted about the axes of the actin filaments, which should have chirally asymmetric torsional compliances due to the filaments' helical microstructure~\cite{egelman1982f,enrique2015actin}.  The resulting ``twirling'' of actin filaments by myosin motors has been directly observed \textit{in vitro} \cite{beausang2008twirling, vilfan2009twirling}.  Furthermore, these torques have been argued to have developmental consequences by contributing to left-right symmetry breaking in the cell cortex  \cite{naganathan2014active, naganathan2016actomyosin}.  However, the intrinsic chirality and torque generation in actin filaments is not captured in any existing cellular scale mechanochemical models of cytoskeletal networks, which currently do not allow for torques or shearing forces due to their one-dimensional filament representations.  Other computational studies also highlight the importance of filament torsion in cytoskeletal assemblies~\cite{ma2018structural,enrique2010origin,yamaoka2010coupling}.  Although these latter studies have implemented models of filament mechanics that include torsional deformations, either the corresponding strain energy functions are overly simple and do not systematically account for all allowable modes of deformation (stretching, bending, shearing, and twisting), or else the models are too computationally expensive to use in network-level simulations of cytoskeletal networks, where collective phenomena involving many filaments are observed~\cite{ma2018structural,enrique2010origin,yamaoka2010coupling}.  Further efforts in modeling actomyosin networks described in Refs. \citenum{cyron2013micromechanical,muller2015resolution} do account for the finite width of actin filaments, but they are limited by large computational expense, an overly simplified set of possible chemical reactions (which excludes active myosin motor walking and filament polymerization), and no option for binding of cross-linkers to the filament surface rather than its backbone.  As a result, the question remains open of how one can incorporate all allowable mechanical deformations of a finite-width filament network in a highly efficient way, so that the model can be used in cellular scale mechanochemical simulation packages such as MEDYAN~\cite{popov2016medyan}.

Here, we introduce a set of options for efficiently modeling a semi-flexible filament having a small but finite width.  The physical background used in these models is the Cosserat theory of elastic rods, which allows for six degrees of freedom at every point on the filament's backbone~\cite{cosserat1909theorie,rubin2000cosserat}.  A key feature of these modeling approaches is that they use only a small, discrete set of model parameters yet describe a smooth filament shape having all allowable degrees of freedom in the Cosserat theory.  We present two main approaches: one in which polynomial spline functions are used to describe the positional and orientational degrees of freedom, and one in which we use the geodesic curve in the space of the orientational degrees of freedom.  We find that in the latter model one can calculate the strain energy function without resorting to a small-angle expansion, so it can describe arbitrarily large filament deformations without systematic error. In Section~\ref{secmethods}, we first introduce the relevant ingredients from the Cosserat theory used in our work, after which we introduce two new filament models and describe how to calculate the strain energy function in each. In Section~\ref{secresults}, we then validate these models by comparing them with computationally expensive but higher resolution Cosserat filament.~\cite{gazzola2018forward} Finally, we apply our new method to illustrate how chiral torsional compliance in a finite-width filament can propagate up a spatial scale to induce broken chiral symmetry in the structure of a cross-linked bundle.  

\section{Methods}\label{Methods}
\label{secmethods}
Thin rods (or filaments) are characterized by large aspect ratios, allowing for an effectively one-dimensional continuum mechanical description where position in the rod is specified with a single variable~\cite{antman2005nonlinear,o2017modeling}.  Several nonlinear theories of thin rod mechanics have been developed which differ from each other in the allowed types of deformations.  The Cosserat theory \cite{cosserat1909theorie,rubin2000cosserat} generalizes the Kirchoff theory \cite{kirchhoff1859ueber,dill1992kirchhoff} by allowing for transverse shearing and axial extension deformations.  Here we build on the Cosserat theory, which has recently been used to develop expressive mathematical models of thin rod dynamics that capture a wide range of observed nonlinear filament behaviors~\cite{bergou2008discrete,gazzola2018forward,zhang2019modeling}.  We next give a brief account of aspects of the Cosserat theory relevant to our model, after which we describe our new variational treatment.

\subsection{Background of Cosserat theory}\label{Background of Cosserat theory}

\begin{figure}
	\begin{center}
		\includegraphics[width=\textwidth]{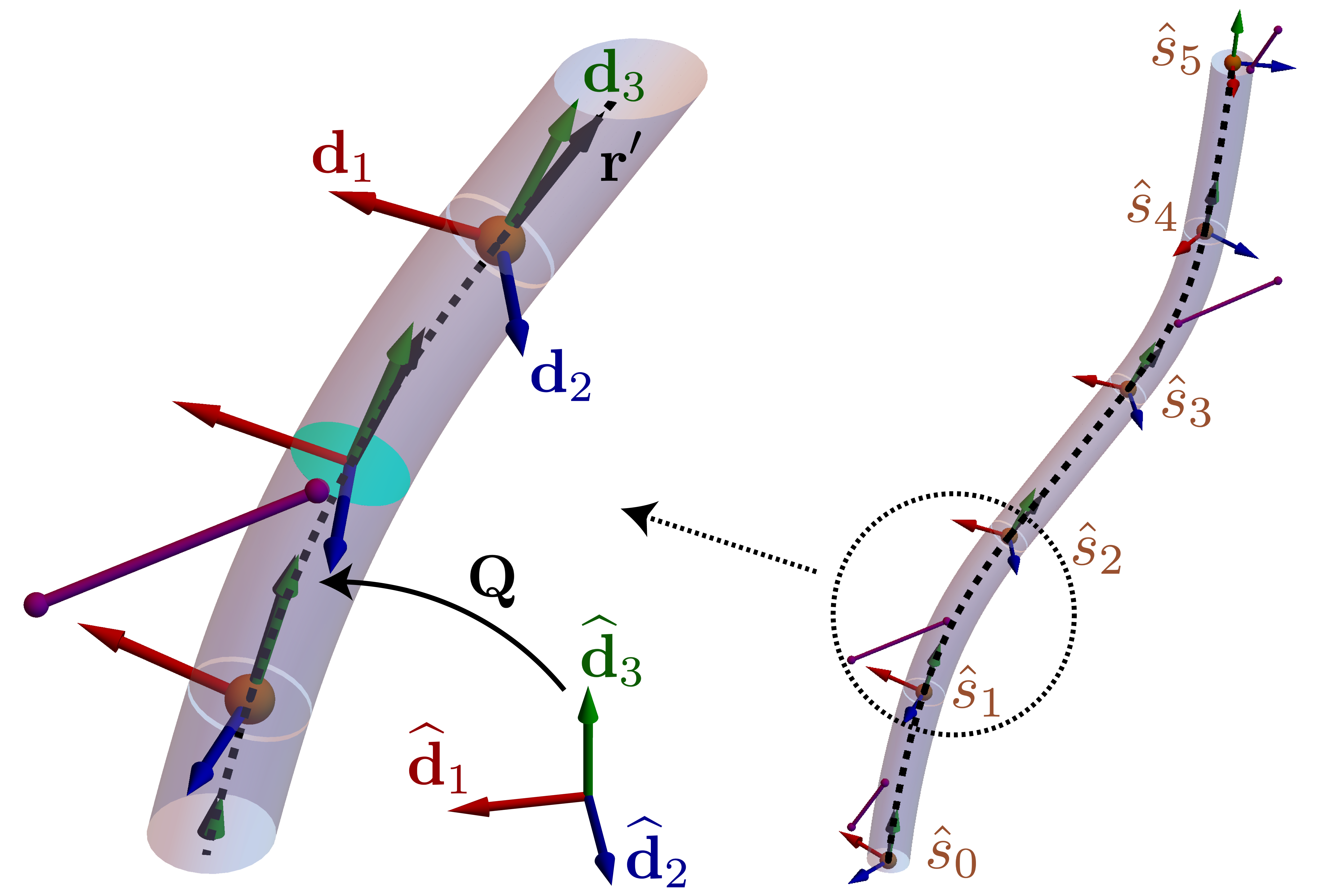}
		\caption{The quantities used to specify a thin rod's configuration in the Cosserat theory are illustrated.  A deformed rod, to whose surface linkers shown in purple are attached, has a backbone $\bm{r}(\hat{s})$ shown as a black dashed curve, and a local triad of directors $\bm{d}_\alpha(\hat{s})$ shown as red, blue, and green arrows.  The matrix $\bm{D}(\hat{s}) = \left(\bm{d}_1(\hat{s}), \ \bm{d}_2(\hat{s}), \ \bm{d}_3(\hat{s}) \right)$ is formed from these column vectors. The rotation tensor $\bm{Q}(\hat{s})$ rotates the reference director triad $\widehat{\bm{d}}_\alpha$ to the current director triad $\bm{d}_\alpha(\hat{s})$.  In the Cosserat theory the local tangent of the backbone $\mathrm{d}\bm{r}/\mathrm{d}\hat{s}$, shown as a black arrow, can differ from the the local $\bm{d}_3(\hat{s})$ director to allow for shear.  The rod is discretized into segments which have endpoints defined by the knot-point coordinates $\hat{s}_i$, shown as orange spheres.  In this visualization there are $5$ segments and $6$ knot points.  The left panel displays a blow-up of the right panel, where the filament's cross-section is visualized as a cyan circle.  Although the director triads are only visualized here at certain discrete positions along the filament, they can be found at any position given the continuous parameterizations used in the variational models.}
		\label{CosseratNotation}
	\end{center}
\end{figure}

In the Cosserat theory, a filament is mathematically described by a directed curve \cite{cosserat1909theorie,cohen1966non}.  This consists of backbone curve $\bm{r}(\hat{s}) \in {\mathbb{R}^3}$, where $\hat{s} \in [0,\hat{L}]$ is the reference arc-length coordinate, and an orthonormal triad of directors $\bm{D}(\hat{s}) = \left(\bm{d}_1(\hat{s}), \ \bm{d}_2(\hat{s}), \ \bm{d}_3(\hat{s}) \right)$. In our notation the caret hat symbol denotes a variable of the reference configuration.  The column vectors $\bm{d}_\alpha(\hat{s}) \in {\mathbb{R}^3}\!$, where $\alpha=1,2,3$ indexes the Cartesian components, specify the orientation of the rod's cross-section at $\hat{s}$ such that $\bm{d}_3(\hat{s})$ is normal to the cross-section and $\bm{d}_1(\hat{s})$ and $\bm{d}_2(\hat{s})$ span the cross-section and define its twist about $\bm{d}_3(\hat{s})$.  These vectors generally differ from the Frenet-Serret frame comprising the tangent, normal, and binormal vectors of $\bm{r}(\hat{s})$~\cite{kreyszig1968introduction}.  We distinguish between the current configuration of the rod, described by $\bm{r}(\hat{s})$ and $\bm{D}(\hat{s})$, and the reference (un-deformed) configuration, described by a reference curve $\hat{\bm{r}}(\hat{s})$ and a reference triad $\widehat{\bm{D}}(\hat{s}) = \left(\widehat{\bm{d}}_1(\hat{s}), \ \widehat{\bm{d}}_2(\hat{s}), \widehat{\bm{d}}_3(\hat{s}) \right).$  Dilatation of the rod's backbone length is captured by the scalar quantity 
$
e(\hat{s}) 
= 
\mbox{d}s/\mbox{d}\hat{s}
$, which can be used to change coordinates from the reference arc-length coordinate $\hat{s}$ to the current arc-length coordinate $s$.  In this paper we primarily use the reference coordinate system, although our formulation is equivalent to one using the current coordinate system.  The proper orthogonal rotation $\bm{Q}(\hat{s}) \in {\mathbb{R}^{3\times3}}$ rotates the reference triad into the current one:
\begin{equation}
	\bm{D}(\hat{s}) = \bm{Q}(\hat{s}) 
	\,
	\bm{\widehat{D}}(\hat{s}).
	\label{Qdef}
\end{equation}
In what follows, we will assume $\widehat{\bm{D}}$ is independent of $\hat{s}$ to simplify notation.  The quantities just introduced are illustrated in Figure~\ref{CosseratNotation}.  

From these quantities, six independent components of strain are defined \cite{antman2005nonlinear,o2017modeling}.  Three of these components are encoded in the pseudovector
\begin{equation}
	\boldsymbol{\kappa} 
	= 
	\text{ax} \! \left(\! \bm{Q}^{T} \,
	\frac{\mbox{d}}{\mbox{d}\hat{s}} \bm{Q} \! \right),
	\label{kappadef}
\end{equation}
where the $\text{ax}$ operation returns the pseudovector associated\footnote{The \text{ax} operation acts on a skew symmetric matrix $\bm{S} = \begin{pmatrix}
		0 & -c & b\\
		c & 0 & -a\\
		-b & a & 0
	\end{pmatrix}$ such that $\text{ax}(\bm{S}) = \bm{v} = (a, b, c)^T$.  Its inverse operation, $\text{skew}$, acts on $\bm{v}$ such that $\text{skew}(\bm{v}) = \bm{S}$.  These have the property that for any vector $\bm{x}$, $\bm{S}\bm{x} = \bm{v} \times \bm{x}$. }  
with the skew-symmetric matrix $\bm{Q}^{T} \, \mbox{d} \bm{Q} \! \ / \ \mbox{d}\hat{s}$. The components of $\boldsymbol{\kappa}$ along the reference triad vectors, $\kappa_\alpha = \boldsymbol{\kappa} \cdot \widehat{\bm{d}}_{\alpha}$, measure the two bending strains, $\kappa_{1}$ and $\kappa_{2}$, and single twisting strain, $\kappa_3$, at each arc-length coordinate~$\hat{s}$.  Similarly, the components along $\widehat{\bm{d}}_\alpha$ of the vector
\begin{equation}
	\boldsymbol{\sigma} = \bm{Q}^{T} \frac{\mbox{d}\bm{r}}{\mbox{d}\hat{s}} - \frac{\mbox{d}\widehat{\bm{r}}}{\mbox{d}\hat{s}}
	\label{sigmadef}
\end{equation}
measure the two transverse shearing strains, $\sigma_{1}$ and $\sigma_{2}$, and the single stretching (extensional) strain, $\sigma_{3}$, completing the collection of six strain measures.  

These strain components are used to define the elastic strain energy density $\varepsilon(\hat{s})$ of the rod's configuration in the Cosserat theory \cite{antman2005nonlinear,o2017modeling}.  The filament's total strain energy $E$ is obtained by integrating the density $\varepsilon(\hat{s})$ along the filament's reference arc-length $\hat{L}$:
\begin{equation}
	E = \int_0^{\hat{L}} \! \varepsilon(\hat{s}) \, \mbox{d}\hat{s} \, .
	\label{Eint}
\end{equation}  The most general quadratic expansion of the local energy density $\varepsilon(\hat{s})$ would include all 27 terms of the form $\kappa_\alpha \kappa_\beta$, $\kappa_\alpha \sigma_\beta$, and $\sigma_\alpha \sigma_\beta$, for $\alpha,\beta = 1,2,3$, capturing mechanical couplings between all components of the strain \cite{enrique2010origin,yamaoka2010coupling}.  However, the material symmetries of the rod can significantly reduce the number of terms needed to describe the energy.  An actin filament can be approximately classified as a transversely hemitropic (i.e. spatially-averaged screw-symmetric) rod \cite{healey2002material}.  It can be shown that such a rod has an energy density with the form \cite{antman2005nonlinear}
\begin{align}
	\varepsilon = &\frac{S_{1,1}}{2}\left(\sigma_1^2 +  \sigma_2^2\right) + \frac{B_{1,1}}{2}\left(\kappa_1^2 +  \kappa_2^2\right) + \frac{S_{3,3}}{2}\sigma_3^2
	+ \frac{B_{3,3}}{2}\kappa_3^2 \nonumber \\
	& + C_{1,1}\left(\sigma_1 \kappa_1 + \kappa_2 \sigma_2 \right) + C_{3,3} \sigma_3 \kappa_3 + C_{1,2}\left(\sigma_1 \kappa_2 - \sigma_2 \kappa_1 \right).
	\label{epsilondef}
\end{align}
The coefficients $S_{\alpha,\beta}$, $B_{\alpha,\beta}$, and $C_{\alpha,\beta}$ can be viewed as elements of the parameter matrices $\bm{S}$, $\bm{B}$, and $\bm{C}$ appearing in the general quadratic expression of $\varepsilon$~\cite{o2017modeling}.  They are geometric and material constants (rigidities) that parameterize the energy penalty of the rod in response to various deformations. The top row of terms in equation (\ref{epsilondef}) captures the energy due to each mode of deformation individually.  The bottom row, which would be zero for an isotropic material, captures couplings between these deformations.  The twist-stretch coupling term proportional to $C_{3,3}$ is of particular interest in the case of actin, because it allows for chiral asymmetry in the filament's torsional compliance \cite{healey2002material,smith2008predicting}.  We note that other mechanical coupling terms not included in equation (\ref{epsilondef}), such as a twist-bend coupling term proportional to $(\kappa_1 + \kappa_2)\kappa_3$, can also play a role in the mechanics of short actin filaments for which spatial averaging is a poor approximation \cite{enrique2010origin}.  Any such coupling terms, as well as higher order terms in the expansion of $\varepsilon$, could be easily accommodated by the methodology subsequently presented in this paper, but in our present applications we use equation (\ref{epsilondef}) to describe chiral actin filaments.  

To parameterize the model we need to specify the elements of $\bm{S}$, $\bm{B}$, and $\bm{C}$ which appear in equation (\ref{epsilondef}).  The diagonal elements of $\bm{S}$ and $\bm{B}$ can be expressed in terms of the material properties $E_\text{mod}$ and $G_\text{mod}$, representing the Young's and shear moduli respectively, and the geometric properties $A$, $\bm{I}$, and $\alpha_c$, representing the cross-sectional area, second (polar) moment of inertia tensor, and a constant equal to $4/3$ for circular cross-sections.  We give the details of the parameterization in the Supplementary material, Section \ref{Parameterization}.  The elements of $\bm{C}$ will be treated here as tunable parameters to study the effect of anisotropic mechanical compliance.  We note that the elements of $\bm{C}$ are bounded by the requirement of energy positivity; for instance $|C_{3,3}| < \sqrt{S_{3,3}B_{3,3}}$ \cite{healey2002material}.

\subsection{Variational approach to rod mechanics}\label{Variational approach to rod mechanics}

In one standard modeling approach, the energy $E$ is used to derive equations of motion which when numerically integrated propagate the rod's configuration forward in time \cite{spillmann2007corde,bergou2008discrete,gazzola2018forward}.  Rather than numerically integrating a  differential equation, which requires significant computational effort for large systems, the approach pursued in this paper is to efficiently find approximately equilibrated rod configurations under some external loads.  In addition to the computational acceleration afforded by directly seeking minimized configurations, we are also motivated to pursue this approach because we aim to incorporate a filament model using the Cosserat theory into the simulation platform MEDYAN \cite{popov2016medyan}.  In MEDYAN, the system's dynamics are propagated forward via short bursts of stochastic chemical activity over a reaction-diffusion compartment grid followed by periodic relaxation of the system's mechanical energy; this allows for efficient simulations that include chemical reactions with spatially varying propensities \cite{baras1996reaction,floyd2020gibbs}.  We note that this energy-minimization based approach to dynamics neglects the thermal diffusive motion of the filaments.  The rationale behind this approach is that the ATP-consuming contributions to the system dynamics, coming from myosin motor steps and actin polymerization, significantly outweighs the contributions coming from diffusive motion of the filaments in these far-from-equilibrium systems \cite{popov2016medyan, mizuno2007nonequilibrium, mackintosh2010active}.  As a result, neglecting thermal motion in MEDYAN simulations is not expected to significantly compromise the realism of the behaviors we are interested in, and this claim has been corroborated through several validations of MEDYAN predictions against experimental measurements~\cite{chandrasekaran2019remarkable, komianos2018stochastic, ni2019turnover, li2020tensile, ni2021membrane,floyd2021understanding}.  Under some external loads, such as cross-linkers bound to the filament, the energy $E$ is variationally minimized for some continuous functions $\bm{r}^\star(\hat{s})$ and $\bm{Q}^{\star}(\hat{s})$, where the star denotes the energy-minimized configuration.  This infinite-dimensional functional minimization problem is computationally burdensome, necessitating a more efficient scheme for scalable simulations.  

The crucial approximation underlying our variational approach is similar in spirit to the Rayleigh-Ritz (or Ritz-Galerkin) method, in which an infinite-dimensional eigenvalue problem is converted to a finite-dimensional one via restriction to a finite-dimensional subspace of expansion coefficients for some chosen basis functions \cite{macdonald1933successive,slaughter2012linearized}.  In our method, we assume that both $\bm{r}(\hat{s})$ and $\bm{Q}(\hat{s})$ are of a specified functional form having a discrete set of free parameters $\mathcal{K}$.  We present two options for this: one in which $\bm{r}(\hat{s})$ and the Euler angles parameterizing $\bm{Q}(\hat{s})$ are spline functions of $\hat{s}$, and one in which $\bm{Q}(\hat{s})$ is the geodesic curve in $\mathrm{SO}(3)$ (the three-dimensional rotation group) on a segment of the filament while $\mbox{d}\bm{r}/\mbox{d}\hat{s}$ is parameterized using its components in the local $\bm{d}_\alpha(\hat{s})$ basis.  The details of these functional forms are elaborated below.  The strain energy density $\varepsilon(\hat{s}; \mathcal{K})$ becomes a function of $\hat{s}$ and the parameters $\mathcal{K}$ through its definition in terms of the strain components, Equation \ref{epsilondef}.  The key challenge in this variational approach is evaluating the integral in Equation \ref{Eint} to express the total energy of the filament $E(\mathcal{K})$ as a function of the model parameters.  Once this is done, equilibrated configurations of the the rod under some external loads, whose energy $E_\text{ext}(\mathcal{K})$ is also expressed in terms of the model parameters, are found by minimizing the total energy $E_\text{tot}(\mathcal{K}) = E(\mathcal{K})+E_\text{ext}(\mathcal{K})$ with respect to the elements in $\mathcal{K}$.  This yields the optimized parameters $\mathcal{K}^\star = \arg \min_\mathcal{K} E_\text{tot}(\mathcal{K})$, which determine the optimized configuration $\bm{r}^\star(\hat{s}) = \bm{r}(\hat{s}; \mathcal{K}^\star)$ and $\bm{Q}^{ \star}(\hat{s}) = \bm{Q}(\hat{s};\mathcal{K}^\star)$.  We find that, using the spline representation for $\bm{r}(s;\mathcal{K})$ and the Euler angles of $\bm{Q}(\hat{s};\mathcal{K})$, it is necessary to expand $\varepsilon(\hat{s};\mathcal{K})$ around small values of the Euler angles for the integral in Equation \ref{Eint} to be analytically solvable.  Using the geodesic form for $\bm{Q}$, this approximation does not need to be made.  We next give the details of these two variational methods.

\subsection{Spline-based models}\label{Spline-based models}

Here we describe how to assign a functional form for the rod's configuration, $\bm{r}(\hat{s})$ and $\bm{Q}(\hat{s})$, using spline functions.  We refer to this approach as the ``spline-based'' model.  Commonly used in the field of computer graphics, several spline functions are available such as B-splines, exponential splines, and Hermite splines, which may each have particular advantages depending on the application \cite{shikin1995handbook,prautzsch2002bezier,spath1969exponential,neuman1978uniform}. For the purpose of demonstrating this approach, we use here composite B\'{e}zier curves which are are fairly intuitive and easy to work with, but this method could be straightforwardly extended to use other splines.  The reference arc-length $\hat{L}$ is discretized into $N_\text{k} - 1$ segments whose ends are $N_\text{k}$ knot coordinates.  The knot coordinates are particular values $\hat{s}_i$ of the reference arc-length, and the $i^{th}$ segment has a reference arc-length $\hat{L}_i = \hat{s}_{i+1} - \hat{s}_i$.  This discretization is illustrated in Figure~\ref{CosseratNotation}.

A composite B\'{e}zier curve $\bm{x}(\hat{s})$ is a piecewise function which passes through $N_\text{k}$ knot points $\bm{x}_i$, where $i = 0,\ldots,N_\text{k}-1$.  On the each segment $i$, $\bm{x}(\hat{s})$ is polynomial of order $d$ whose shape is controlled by the control points $\bm{x}_{i,j}$, where $j = 0,\ldots,d-2$ and the double index indicates that $\bm{x}_{i,j}$ is a control point.  Like $N_\text{k}$, the polynomial order $d$ is a hyperparameter controlling the complexity of the model.  We will formulate the model for general values of these hyperparameters, but in our implementations we choose $d$ as $1$ or $2$ and $N_\text{k}$ such that an actin filament segment is $\sim 20 - 100 \ \mbox{nm}$ long.  In Figure~\ref{CosseratNotation}, for instance, we have $N_\text{k} = 6$ which is a typical value used.  The full curve $\bm{x}(\hat{s})$ consists of $N_\text{k}-1$ segments $\bm{x}_i(\hat{s})$, $i = 0,\ldots, N_\text{k} - 2$, such that $\bm{x}(\hat{s}) = \bm{x}_i(\hat{s})$ if $\hat{s}_i \leq \hat{s} < \hat{s}_{i+1}$.  Here the argument of $\bm{x}_i(\hat{s})$ indicates that it is a function rather than a knot point.  The segment curves $\bm{x}_i(\hat{s})$ are reparameterized using the segment variable $q(\hat{s};\hat{s}_{i},\hat{s}_{i+1}) = (\hat{s} - \hat{s}_i)/\hat{L}_i$ which ranges from $0$ to $1$ as $\hat{s}$ increases from $\hat{s}_i$ to $\hat{s}_{i+1}$.  In terms of $q$,  the segment curves are given by the formula
\begin{equation}
	\bm{x}_i(q) = (1-q)^d\bm{x}_i + q^d \bm{x}_{i+1} + \sum_{i=0}^{d-2}B_{j+1}^d(q)\bm{x}_{i,j},
	\label{segmentcurve}
\end{equation}
where the Bernstein polynomials are
\begin{equation}
	B_{j}^d(q) = \binom{d}{j}q^j(1-q)^{d-j}.
	\label{Bernstein}
\end{equation}
If all knot and control points are free, then un-physical cusps can result in the composite curve at the knot points.  To address this, smoothness up to degree $p$ can be enforced through derivative matching conditions 
\begin{equation}
	\lim_{\hat{s} \rightarrow \hat{s}_{i+1}^-}\bm{x}_{i}^{(k)}(\hat{s}) = \lim_{\hat{s} \rightarrow \hat{s}_{i+1}^+}\bm{x}_{i+1}^{(k)}(\hat{s})
	\label{derivmatching}
\end{equation}
where $ i = 0, \ldots N_k - 3, \ k = 1,\ldots,p$ and $\bm{x}_{i}^{(k)}(\hat{s})$ denotes the $k^\text{th}$ derivative with respect to $\hat{s}$ of $\bm{x}_{i}(\hat{s})$.  Choosing $p=d-1$ gives $3(d-1)(N_\text{k}-2)$ equations in $3(N_\text{k} + (N_\text{k}-1)(d-1))$ parameters, leaving $3(N_\text{k} + d -1)$ free.  These free parameters can be taken to be the $N_\text{k}$ knot points $\bm{x}_i$ and the control points on the first segments, $\bm{x}_{0,j}$, $j=0,\ldots,d-2$.  This choice of $p$ provides the greatest amount of smoothness while also allowing the number of free parameters to grow with $N_\text{k}$.  One deficit of this parameterization is that specifying a position on the $i^{th}$ segments requires using parameters from segments $0$ to $i-1$, since the control points on the $i^{th}$ segment are determined from the smoothness constraints involving these previous parameters.  However, for reasonably small values of $N_\text{k}$, up to $\sim 10$, this issue does not significantly impair model performance (we discuss the computational efficiency of these models in the Supplementary material, Section \ref{Timing and accuracy of the models}).

\sloppy
A composite B\'{e}zier curve is used to represent both the backbone curve $\bm{r}(\hat{s})$ as well as the curve $\boldsymbol{\alpha}^\text{Eu}(\hat{s})$, containing the Euler angles parameterizing $\bm{Q}(\hat{s})$, as function of $\hat{s}$.  The vector $\boldsymbol{\alpha}^\text{Eu}(\hat{s}) = (\phi^\text{Eu}(\hat{s}), \ \theta^\text{Eu}(\hat{s}), \ \psi^\text{Eu}(\hat{s}))^T$ encodes here the 3-2-1 (yaw-pitch-roll) Euler angles of $\bm{Q}$, although other Euler angle conventions could also be used \cite{o2008intermediate}.  For this representation of $\bm{r}(\hat{s})$ and $\bm{Q}(\hat{s})$, the model parameters are $\mathcal{K} = \{ \bm{r}_{0,0} ,\ldots ,\bm{r}_{0,d_r-2}, \bm{r}_{0}, \ldots ,\bm{r}_{N_\text{k} - 1},\boldsymbol{\alpha}^\text{Eu}_{0,0} ,\ldots ,\boldsymbol{\alpha}^\text{Eu}_{0,d_\alpha-2}, \boldsymbol{\alpha}^\text{Eu}_{0}, \ldots ,\boldsymbol{\alpha}^\text{Eu}_{N_\text{k} - 1} \}$ where $d_r$ and $d_\alpha$ are the orders of the composite B\'{e}zier curves for the backbone and Euler angles respectively.  Using the above definitions of the strain components which enter into Equation \ref{epsilondef}, it is straightforward to write the strain energy density $\varepsilon(\hat{s};\mathcal{K})$ using these spline parameters.  However, to find the integrated energy $E(\mathcal{K})$, it is necessary due to the intractability of analytically integrating $\varepsilon(\hat{s};\mathcal{K})$ to make a small-angle approximation to $\bm{Q}(\boldsymbol{\alpha}^\text{Eu}) \approx \bm{Q}_\text{approx.}(\boldsymbol{\alpha}^\text{Eu};m)$.  The order $m$ of the small-angle expansion is an additional hyperparameter of the model.  Using $\bm{Q}_\text{approx.}$ in place of $\bm{Q}$ in the definition of the strain components, the approximate strain energy density $\varepsilon_\text{approx.}(\hat{s};\mathcal{K})$ becomes a polynomial in $\hat{s}$, that is 
\begin{equation}
	\varepsilon_\text{approx.}(\hat{s};\mathcal{K}) = \sum_{k} \varepsilon_k(\mathcal{K}) \hat{s}^k,
	\label{splineapproxenergy}
\end{equation}
which may therefore be easily integrated to give the approximate energy of the filament $E_\text{approx.}(\mathcal{K})$.  We implemented a routine to calculate $E_\text{approx.}(\mathcal{K})$ using the computer algebra system Mathematica \cite{mathematica}.  The details of this calculation are tedious (though straightforward) and do not provide additional insight, so we do not present them here.  They can be found in the accompanying Mathematica notebooks.  

\subsection{Geodesic models}\label{Geodesic models}
As discussed in Section~\ref{secresults}, the small-angle approximation used to obtain an analytical expression for the integral of the energy density in the spline-based model can lead to biased filament configurations that are highly inaccurate when the deformations are large.  To address this issue, we next present a so-called ``geodesic'' model which avoids making the small-angle approximation and produces approximately correct filament configurations even for large deformations.  

In the geodesic model, we adopt the axis-angle parameterization for the rotation tensor $\bm{Q}$, rather than the Euler angle parameterization used above.  In the axis-angle parameterization $\bm{Q}(\bm{u}, \theta)$ represents a rotation about the unit vector $\bm{u}$ by the angle $\theta$.  The Rodrigues formula expresses the tensor in terms of $\bm{u}$ and $\theta$ as
\begin{equation}
	\bm{Q}(\bm{u}, \theta) = \cos(\theta)\left(\bm{E} - \bm{u} \otimes \bm{u} \right) + \sin(\theta)\text{skew}( \bm{u}) + \bm{u} \otimes \bm{u},
	\label{rodrigues}
\end{equation}
where $\bm{E}$ is the $3\times3$ identity matrix, the $\text{skew}$ operation returns the skew-symmetric matrix matrix associated with the unit vector $\bm{u}$ (and is the inverse of the $\text{ax}$ operation), and $\otimes$ denotes the outer (dyadic) product \cite{o2017modeling}.  $\bm{Q}(\bm{u}, \theta)$ can also be represented using matrix exponentiation as 
\begin{equation}
	\bm{Q}(\bm{u}, \theta) = \exp\left( \text{skew}(\theta\bm{u})\right).
	\label{exprot}
\end{equation}
This representation makes evident the connection between the orthogonal tensor $\bm{Q}$, a member of the Lie group $\mathrm{SO}(3)$, and the skew-symmetric tensor $\text{skew}(\theta \bm{u})$, a member of the associated Lie algebra $\mathfrak{so}(3)$ \cite{jeevanjee2011introduction}.  The transpose $\bm{Q}^T(\bm{u}, \theta)$ can be obtained as $\bm{Q}^T(\bm{u}, \theta) = \bm{Q}(\bm{u}, -\theta)$. 

We again use $N_\text{k}$ knot coordinates to discretize the filament into $N_\text{k}-1$ segments.  At every knot coordinate $\hat{s}_i$ a rotation tensor $\bm{Q}_i$ is parameterized with free model parameters $\bm{u}^\text{Ax}_i$ and $\theta^\text{Ax}_i$.  To enforce normalization, we represent $\bm{u}^\text{Ax}_i$ in polar coordinates using the polar and azimuthal angles, $\beta^\text{Ax}_i$ and $\gamma^\text{Ax}_i$ respectively.  Thus the collection of angles $\theta^\text{Ax}_i, \ \beta^\text{Ax}_i$, and $\gamma^\text{Ax}_i$ parameterize $\bm{Q}_i$ at $\hat{s}_i$.  On the $i^{th}$ segment (where $\hat{s}_i \leq \hat{s} < \hat{s}_{i+1}$), the rotation tensor $\bm{Q}_i(q)$ is taken to be the geodesic curve on the manifold $\mathrm{SO}(3)$ which connects the two tensors $\bm{Q}_i$ and $\bm{Q}_{i+1}$, where $q(\hat{s};\hat{s}_{i},\hat{s}_{i+1}) = (\hat{s} - \hat{s}_i)/\hat{L}_i$ is the local segment variable as in the spline-based model.  The geodesic curve depends on the metric $D$ used to define distances in $\mathrm{SO}(3)$, and we use the metric
\begin{equation}
	D(\bm{Q}_A, \bm{Q}_B) = \lvert \theta_{A,B} \rvert,
	\label{metric}
\end{equation} 
where $\theta_{A,B}$ is the angle in the axis-angle parameterization of the tensor $\bm{Q}_B \bm{Q}_A^T$ rotating $\bm{Q}_A$ to $\bm{Q}_B$ \cite{huynh2009metrics}.  It can be shown that the geodesic curve $\bm{Q}_i(q)$ connecting $\bm{Q}_i$ and $\bm{Q}_{i+1}$ using this metric is
\begin{equation}
	\bm{Q}_i(q) = \exp\left(q \ln \left( \bm{Q}_{i+1} \bm{Q}^T_i \right)\right)\bm{Q}_i,
	\label{geodesiccurve}
\end{equation}
where $\ln$ is the matrix logarithm \cite{park1995distance,park1997smooth}.  As described in the Supplementary material, Section \ref{Functions in the goedesic models} this curve can then be expressed in terms of the free model parameters $\theta^\text{Ax}_i, \ \beta^\text{Ax}_i, \ \gamma^\text{Ax}_i, \ \theta^\text{Ax}_{i+1}, \ \beta^\text{Ax}_{i+1}$, and  $\gamma^\text{Ax}_{i+1}$.  The global tensor curve $\bm{Q}(\hat{s})$ is given piecewise by $\bm{Q}_i(\hat{s})$ on the $N_\text{k}-1$ segments.

To represent the backbone curve $\bm{r}(\hat{s})$ in the geodesic model, we write its derivative with respect to $\hat{s}$ in the local director triad basis $\bm{d}_\alpha$:
\begin{equation}
	\frac{\mbox{d} \bm{r}(\hat{s})}{\mbox{d} \hat{s}} = \zeta_\alpha(\hat{s}) \bm{d}_\alpha(\hat{s})
	\label{rprimegeodesic}
\end{equation}
where summation over repeated indices is implied.  Here we treat the components $\zeta_\alpha(\hat{s})$ as constants on each segment, i.e. $\zeta_\alpha(\hat{s}) = \zeta_{i,\alpha}$ for $\hat{s}_i \leq \hat{s} < \hat{s}_{i+1}$, although this assumption could be relaxed.  The piecewise constant components $\zeta_{i,\alpha}$ are additional free model parameters.  We refer to the model where $\zeta_{i,\alpha}$ are all independent as the ``geodesic Cosserat'' (GC) model.  We can also optionally set $\zeta_{i,1} = \zeta_{i,2} = 0$ on all segments, implying that $\mbox{d}\bm{r}/\mbox{d}\hat{s}$ is everywhere parallel to $\bm{d}_3(\hat{s})$ such that there is zero shear on the filament. The filament is still extensible, since $\zeta_{i,3} \neq 0$, so this model is referred to as the ``geodesic extensible Kirchoff'' (GEK) model.  To obtain the backbone curve $\bm{r}(\hat{s})$, we integrate $\mbox{d}\bm{r}/\mbox{d}\hat{s}$ from the minus-end position of the filament $\bm{r}_0 = \bm{r}(\hat{s} = 0)$, as shown in Equation \ref{eqrGEK} in the Supplementary material, Section \ref{Functions in the goedesic models}.  The initial point $\bm{r}_0$ is the final free parameter in the GC and GEK models.  We note that the filament energy $E(\mathcal{K})$ will not depend on $\bm{r}_0$ due to translation invariance, but external potentials $E_\text{ext}(\mathcal{K})$ such as cross-linkers bound to the filament will depend on $\bm{r}_0$.  For the GC model, there are $6 N_\text{k}$ elements in $\mathcal{K}$, while for the GEK model there $4 N_\text{k} + 2$.

A significant benefit of the geodesic parameterization of $\bm{r}(\hat{s})$ and $\bm{Q}(\hat{s})$ is that it allows the energy density to be analytically integrated along the length of the filament to give an exact expression for the total filament energy $E(\mathcal{K})$.  We describe the derivation of this expression in the Supplementary material, Section \ref{Energies in the geodesic models}.  As an intuitive picture, the geodesic curves used in this model can be thought of as representing a ``linear'' interpolation between rotation tensors in their natural mathematical space, and they are therefore expected to be a useful tool for parameterizing how several free rotation tensors are connected together.

\subsection{MEDYAN model}\label{MEDYAN filament model}
For comparison, we describe here the original zero-width mechanical model used in the simulation platform MEDYAN \cite{popov2016medyan}.  This model has no allowed shearing or twisting, although excluded volume repulsion is included between filaments using a finite effective filament radius \cite{floyd2021segmental}.  The energy in this model does allow for stretching and bending.  The filament is again discretized into $N_\text{k}$ knot coordinates and $N_\text{k} - 1$ segments.  Each segment is a straight line with a current length $L_i = s_{i+1} - s_i$ and a reference length $\hat{L}_i = \hat{s}_{i+1} - \hat{s}_i$.  The stretching energy on each segment is a quadratic function of these lengths:
\begin{equation}
	E_i^\text{stretch} = \frac{S_{3,3}}{2 \hat{L}_i}\left(L_i - \hat{L}_i \right)^2.
	\label{MEDstretch}
\end{equation}
The segment acts like a spring with spring constant $S_{3,3} / \hat{L}_i$.  At each internal knot coordinate there is a bending potential involving the angle $\theta^\text{MED}_{i,i+1}$ between the $i^{th}$ and $(i+1)^{th}$ segment:
\begin{equation}
	E_i^\text{bend} = \frac{B_{1,1}}{\hat{L}_i}\left(1 - \cos\left(\theta^\text{MED}_{i,i+1}\right)  \right),
	\label{MEDbend}
\end{equation}
where $i = 1, \ldots, N_\text{k}-2$.  We show in the Supplementary material, Section \ref{Energies in the geodesic models} that this expression for the bending energy can be obtained as a special case of the GC model bending energy.  In the MEDYAN model, the knot points $\bm{r}_i$ are the only free parameters, and the backbone curve $\bm{r}(\hat{s})$ is a linear interpolation between these points.  There are thus $3 N_\text{k}$ elements in $\mathcal{K}$ for this model.  

\subsection{Dynamical model}\label{Dynamical model}
We also briefly describe for comparison the dynamical model of a filament developed by Gazzola et al. in Ref. \citenum{gazzola2018forward}.  Rather than directly seeking equilibrated configurations of the filament, the dynamical approach propagates the filament's configuration forward in time using discretized equations of motion based on the forces and torques in the filament.  Propagating the configuration forward for long times with dissipation will cause the filament to converge to its equilibrated configuration under some external loads.  The filament in this model is discretized into $N_\text{n} - 1$ linear segments which, when $N_\text{n}$ is large, allows for a good approximation to any arbitrary filament backbone configuration.  In our usage here, we take $N_\text{n} \gg N_\text{k}$, so there are far fewer degrees of freedom in the variational models than in the dynamical model.  The backbone $\bm{r}_i(t)$ is specified at each of the $N_\text{n}$ coordinates, and the rotation tensor $\bm{Q}_i(t)$ is specified on each of the $N_\text{n} - 1$ segments.  These quantities are updated in discrete time steps $\delta t$ using a second-order velocity Verlet integrator scheme.  The equations of motion correspond to an isotropic energy function given by the top row of terms in Equation \ref{epsilondef}.  For fine spatial and temporal discretization this model has been shown to be very expressive, capturing a range of realistic filament behaviors, and we use it in this paper as the ``ground truth'' to which our computationally accelerated variational models can be compared to assess their accuracy.  We refer the reader to Ref. \citenum{gazzola2018forward} for details of this model, and to the follow-up paper, Ref. \citenum{gunaratne2022stretching}, in which extensions to this dynamical model are developed to treat multiple spatial scales simultaneously.  

\subsection{Binding to surface}\label{Binding to surface}
\begin{figure}[H]
	\begin{center}
		\includegraphics[width=13 cm]{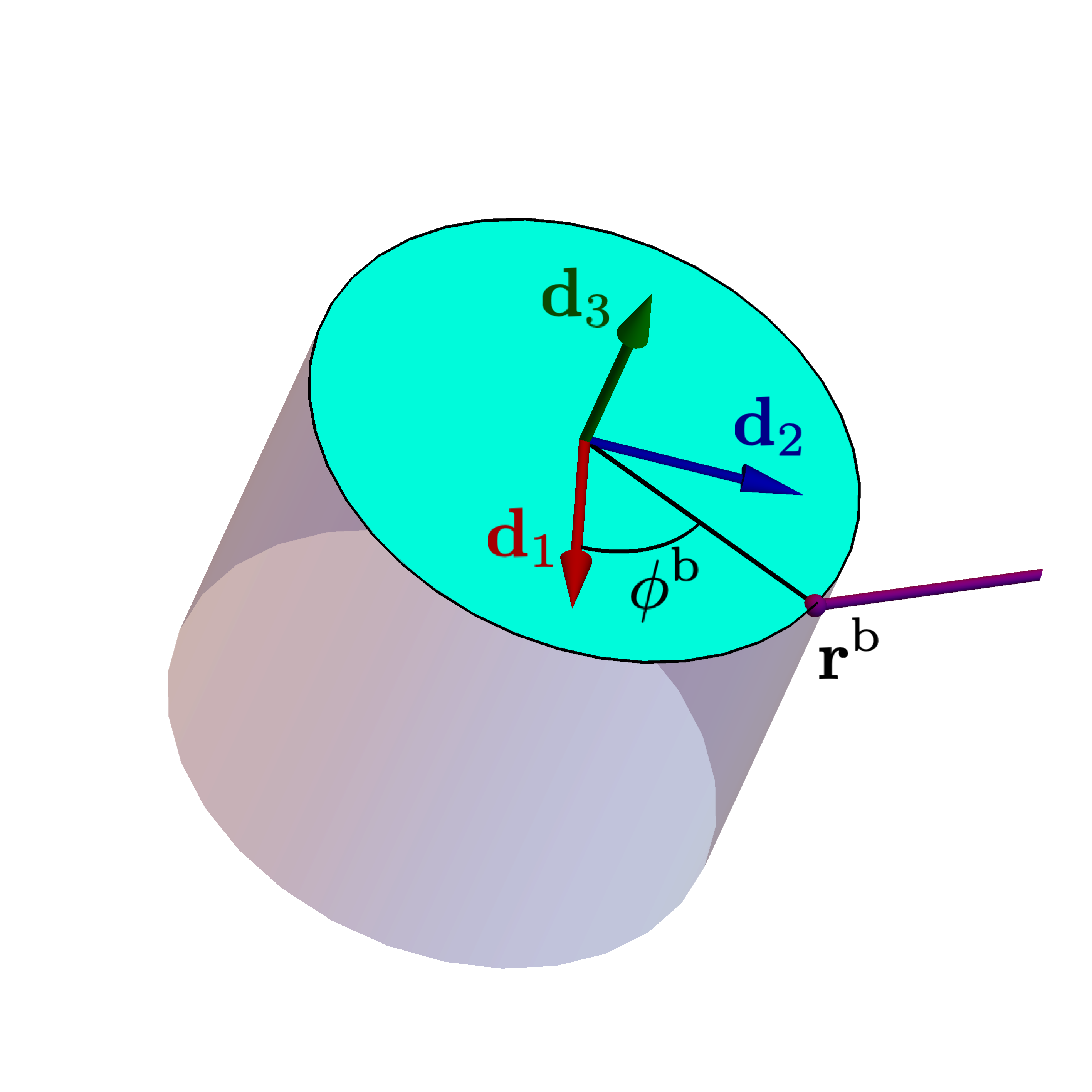}
		\caption{The cross-section of a filament with a bound cross-linker is shown.  The cross-linker is visualized in purple and the local director triad is visualized as a set of colored vectors.  The angle $\phi^\text{b}$ and bound position $\bm{r}^\text{b}$ appearing in Equation \ref{eqcrosssection} are also illustrated. }
		\label{CrossSection}
	\end{center}
\end{figure}

In one-dimensional filament models like the original MEDYAN model, external loads such as bound cross-linkers on the filament are attached directly to the filament backbone.  In the new models presented above, the filament has a finite width and external loads may attach to the surface of the filament, exerting shearing and twisting forces.  This introduces an extra degree of freedom at the attached arc-length coordinate $\hat{s}^\text{b}$ corresponding to the position on the perimeter of the filament's cross-section at $\hat{s}^\text{b}$ to which the load is attached.  This cross-section is spanned by the vectors $\bm{d}_1(\hat{s}^\text{b})$ and $\bm{d}_2(\hat{s}^\text{b})$, and the one-dimensional position on the perimeter of the cross-section can be parameterized by the polar angle $\phi^\text{b}$ with respect to the local $\bm{d}_1(\hat{s}^\text{b})$ axis.  For a circular cross section, the position of the attached load is given by 
\begin{equation}
	\bm{r}^\text{b} = \bm{r}(\hat{s}^\text{b}) + R \left(\cos \left(\phi^\text{b}\right) \bm{d}_1(\hat{s}^\text{b}) + \sin \left(\phi^\text{b}\right) \bm{d}_2(\hat{s}^\text{b}) \right), 
	\label{eqcrosssection}
\end{equation}
where $R$ is the filament radius.  This position therefore couples not just to the backbone $\bm{r}(\hat{s}^\text{b})$ but also to the local rotation tensor $\bm{Q}(\hat{s}^\text{b})$ through $\bm{d}_\alpha(\hat{s}) = \bm{Q}(\hat{s}) \widehat{\bm{d}}_\alpha$.  This binding of an attached linker to the local perimeter of the filament cross-section is illustrated in Figure~\ref{CrossSection}.

By allowing for cross-linkers to bind to the surface of a filament rather than its backbone, an extra degree of freedom $\phi^\text{b}$ is introduced.  We note that during a simulated binding event, this degree of freedom could be chosen in several ways, which we describe in the Supplementary material. For example, to provide additional biological realism a modeller could fix $\phi^\text{b}$ for the possible binding sites to lie along a helix which wraps around the filament. In this way, the helical microstructure of actin could be encoded into the available binding sites.  

\section{Results}\label{Results}
\label{secresults}

Here we present both validation and application results using the models described above.  Model validation is done by comparing the equilibrated filament configurations for different test cases to the finely discretized dynamical model of Gazzola et al., which we take to be the ground truth \cite{gazzola2018forward}.  The results indicate that under small applied loads all models agree well, but for larger applied loads the geodesic models, which avoid the small-angle approximation, perform significantly better than the spline-based models which exhibit systematic bias.  In the Supplementary material, Section \ref{Euler buckling study}, we show that the geodesic models also better reproduce the theoretical predictions of Euler buckling compared to the spline-based models.  The original zero-width MEDYAN model also produces systematically unbiased rod configurations under large deformations but does not allow for finite filament widths.  To demonstrate a novel application of the new finite-width models, we simulate a ``chiral bundle,'' a group of seven filaments with non-zero chiral coupling rigidity $C_{3,3}$ interconnected by surface-bound cross-linkers.  Pulling vertically on the bundle induces a twist of each filament which, through the attached cross-linkers, causes the entire bundle structure to adopt a twisted configuration.  Such a structure may be relevant to the physiological functioning of actin stress fibers in eukaryotic cells and also demonstrates the possibility of intrinsic filament chirality propagating up a spatial scale to break chiral symmetry in the structure of a cross-linked bundle \cite{tojkander2012actin}.

\subsection{Model comparisons}\label{Model comparisons}
To assess the accuracy of the spline-based and geodesic models introduced above, we compare their equilibrated configurations to the steady state configuration of the finely-discretized dynamical model with dissipation.  Five models are compared against the dynamical model: the B\'{e}zier spline-based model with a first order small-angle expansion (``B, $ m=1$''), the B\'{e}zier spline-based model with a second order small-angle expansion (``B, $m=2$''), the geodesic extensible Kirchoff model (GEK), the geodesic Cosserat model (GC), and the original MEDYAN model.  To quantify the difference between the equilibrated configurations, we use two similarity measures.  The first, $C_r$, measures the root-mean-squared distance between the backbone curves of two filaments $A$ and $B$:
\begin{equation}
	C_r\left(\bm{r}^A(\hat{s}), \bm{r}^B(\hat{s})\right) = \left(\hat{L}^{-1} \int_0^{\hat{L}}\lvert\lvert \bm{r}^A(\hat{s}) - \bm{r}^B(\hat{s})\rvert\rvert^2 \mbox{d} \hat{s} \right)^{1/2},
	\label{Crdef}
\end{equation}
where $\lvert\lvert \cdot \rvert\rvert$ denotes the vector norm.  The second similarity measure, $C_d$, is introduced to measure the average difference in the vectors $\bm{d}_3^A(\hat{s})$ and $\bm{d}_3^B(\hat{s})$ along the reference arc-length:
\begin{equation}
	C_d\left(\bm{d}_3^A(\hat{s}), \bm{d}_3^B(\hat{s})\right) = \hat{L}^{-1} \int_0^{\hat{L}}\left(1 - \bm{d}_3^A(\hat{s}) \cdot \bm{d}_3^B(\hat{s}) \right) \mbox{d} \hat{s}.
	\label{Cddef}
\end{equation}
Both $C_r$ and $C_d$ are evaluated numerically using a large number of sample points.

\begin{figure}[H]
	\begin{center}
		\includegraphics[width=9.5 cm]{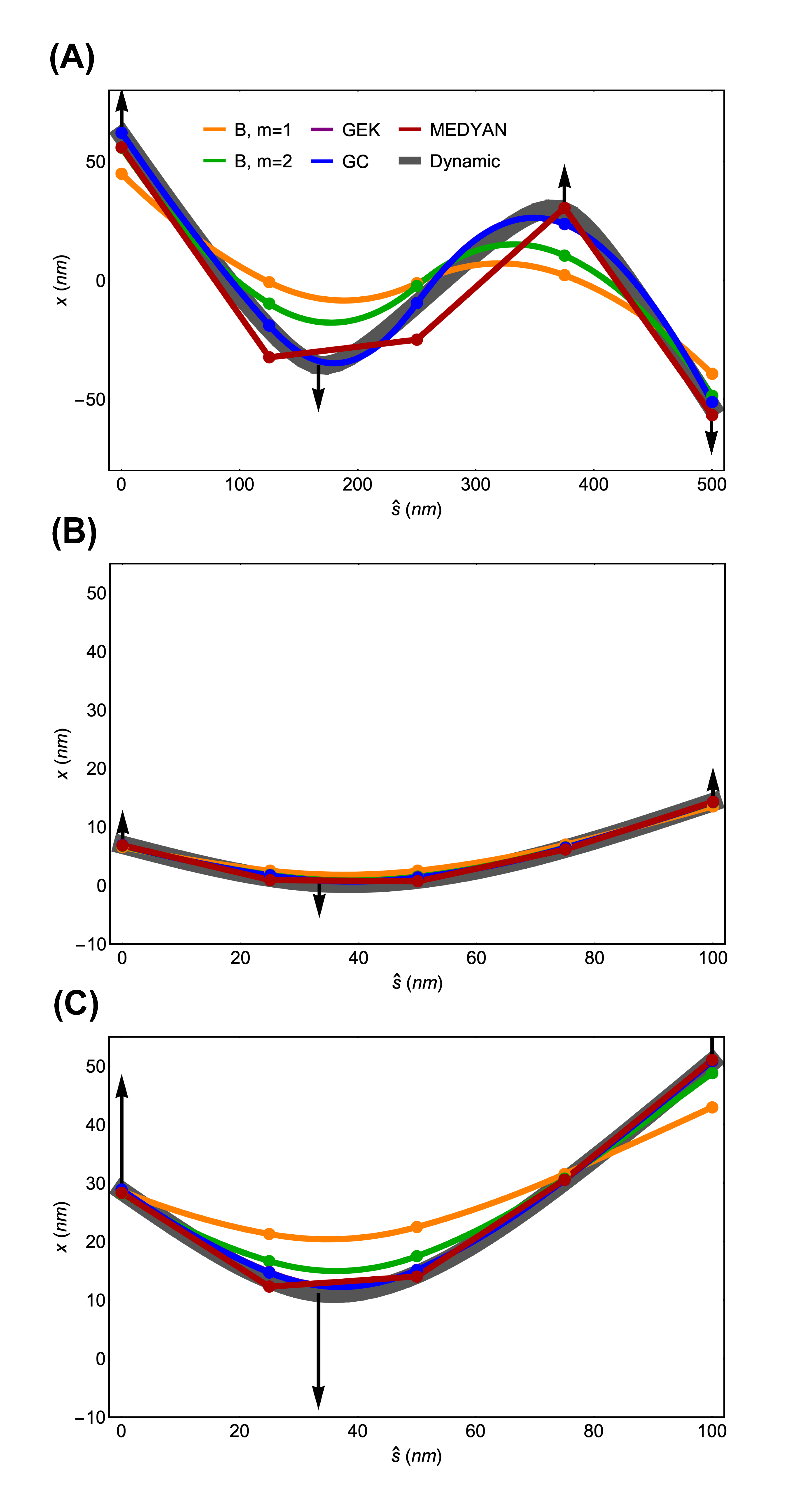}
		\caption{A comparison of the variational and dynamical models is shown.  (A)  The equilibrated backbone curve, projected onto the $x$-axis, for the five variational models and the dynamical model are shown in the bottom plot, colored according to the legend.  This panel corresponds to the first test case described in the main text.  The five knot points of the variational models are shown as solid circles through which the curves pass.  The black arrows attached to the dynamical solution's curve indicate the attachment point and direction of the bound cross-linkers, and the arrow lengths are proportional to the $x$ coordinate of the cross-linkers' other ends. (B) (resp. (C)) This panel is the same as panel (A), except it applies to the second (resp. third) test case described in the main text. We note that the GEK model curve is closely matched by the GC model curve and is hidden behind it in these graphs. }
		\label{CompositeComparison}
	\end{center}
\end{figure}

Three test cases, labeled (A), (B) and (C) in Figure~\ref{CompositeComparison} were used to assess the models' accuracy.  The deformation in each test case is in a 2D plane so that the configurations can be easily visualized.  In test case (A), a $500$ nm-long filament is pulled by four springs in opposing directions, while in test cases (B) and (C), a $100$ nm long filament is pulled by three springs.  Test cases (B) and (C) are distinguished by the strength of the pulling such that the springs in test case (C) are much more stretched than those in (B), causing a greater filament deformation.  The filaments in each test case are modeleled as isotropic (with the coupling matrix elements $C_{\alpha,\beta}$ in Equation \ref{epsilondef} set to zero), and the springs attach directly to the filament backbone.  This is done to allow comparison with the original MEDYAN model and the dynamical model implementations, which do not currently support anisotropic filaments or surface-bound cross-linkers.  The details of the set up for these test cases are described in the Supplementary material, Section \ref{Description of test cases}.

For the three test cases, we generally observed that the closest agreement with the dynamical solution was obtained by the geodesic models, however for the small deformations in test case (B) all models agree well with each other (see Table \ref{Distances}).  This indicates the sufficiency of the small-angle approximation for small applied loads.  For large loads, the spline-based models have systematically smaller deformations than the geodesic and dynamical models, which is a major shortcoming.  The original MEDYAN model does not exhibit this systematic error though it is less precise due to its linear segment shapes.  The difference between the GEK and GC models is negligible for all test cases, resulting from the high shearing modulus of actin; the extra degrees of freedom in the GC model should still be useful for modeling other types of filaments.  Both geodesic models agree very well with the dynamical solution, with only slight differences in shape even for large applied loads.  Minimizing the geodesic models to obtain the equilibrated configurations takes on the order of seconds of computational time, however, whereas propagating the finely discretized dynamical model until it is equilibrated takes on the order of days.  We display in the Supplementary material, Section \ref{Strain profiles} the profile of shearing, extensional, bending, and twisting strain for the variational models along the length of the equilibrated filament for the third test case.  We also show in the Supplementary material, Section \ref{Timing and accuracy of the models} the computational timing of the variational models along with their accuracy as $N_\text{k}$ is varied.  Finally, we show in Table \ref{TableEnergies} the equilibrated filament and spring energies for each test case.  We see that in each case the dynamical model achieves the smallest total energy of all models, which we might expect due to its comparative lack of restrictions on the filament configuration.  However, the agreement in energy between the geodesic models and the dynamical model is excellent.  Although in these test cases the difference between the GEK and GC models is small, we expect that in the context of real cytoskeletal networks, where molecular motors and branching molecules bind to the surfaces of filaments, the shearing degrees of freedom in the GC will be important.  Such bound molecules can produce localized shearing forces which would not be resolved in the GEK model.

\begin{table}[H]
	\begin{center}
		\begin{tabularx}{0.85\textwidth}{  l l r r r r r }
			\hline
			\textbf{Case} & \textbf{Metric} & $\bm{B, \ m=1}$ & $\bm{B, \ m=2}$ & \textbf{GEK} & \textbf{GC} & \textbf{MEDYAN} \\ 
			\hline
			(A)  & $C_r$ (nm) & $18.40$ & $13.05$ & $3.89$ & $3.86$ & $9.16$ \\
			& $C_d$ & $0.041$ & $0.022$ & $0.008$ & $0.008$ & None \\ 
			\hline
			(B)  & $C_r$ (nm) & $1.06$ & $0.39$ & $0.30$ & $0.28$ & $0.23$  \\
			& $C_d$  & $0.001$ & $0.000$ & $0.000$ & $0.000$ & None \\ 
			\hline
			(C)  & $C_r$ (nm) & $6.31$ & $2.69$ & $0.84$ & $0.80$ & $0.93$ \\
			& $C_d$ & $0.036$ & $0.011$ & $0.000$ & $0.001$ & None \\ 
			\hline
		\end{tabularx}
		\caption{The difference metrics $C_r$ and $C_d$ between each model and the dynamical model are reported for the three test cases.  The label of each test case matches the panel of Figure~\ref{CompositeComparison} where that case is visualized.}
		\label{Distances}
	\end{center}
\end{table}

\begin{table}[H]
	\begin{center}
		\begin{tabularx}{\textwidth}{  l l r r r r r r }
			\hline
			\textbf{Case} & \textbf{Energy} & $\bm{B, \ m=1}$ & $\bm{B, \ m=2}$ & \textbf{GEK} & \textbf{GC} & \textbf{MEDYAN}& \textbf{Dynamic} \\ 
			\hline
			& $E$ & $6,458$ & $10,179$ & $6,351$ & $6,365$ & $3,986$ & $6,806$\\
			(A)& $E_\text{ext}$ & $104,842$ &  $81,320$ & $66,038$ & $65,978$ & $65,531$ & $60,024$ \\ 
			& $E_\text{tot}$ & $111,300$ & $91,499$ & $72,389$ & $72,343$ & $69,517$ & $66,830$ \\ 
			\hline
			& $E$ & $712$ & $900$ & $936$ & $940$ & $935$ & $980$ \\
			(B) & $E_\text{ext} $ & $4,423$ & $4,115$ & $4,068$ & $4,056$ & $4,065$ & $3,961$ \\ 
			& $E_\text{tot} $ & $5,135$ & $5,015$ & $5,004$ & $4,996$ & $5,000$ & $4,941$ \\ 
			\hline
			& $E $ & $3,895$ & $6,966$ & $8,689$ & $8,689$ & $8,160$ & $9,099$ \\
			(C) & $E_\text{ext} $ & $95,747$ & $87,669$ & $84,076$ & $83,990$ & $84,792$ & $82,989$\\ 
			& $E_\text{tot} $ & $99,642$ & $94,635$ & $92,765$ & $92,679$ & $92,952$  &  $92,088$\\ 
			\hline
		\end{tabularx}
		\caption{The equilibrated filament energies $E$ (top row for each model), external energies from bound linkers $E_\text{ext}$ (middle row), and total energy $E_\text{tot}$ (bottom row) are reported for the three test cases.  The label of each test case matches the panel of Figure~\ref{CompositeComparison} where that case is visualized.  All units of energy are pN  nm.}
		\label{TableEnergies}
	\end{center}
\end{table}

\subsection{Bundle study}\label{Bundle study}

Here we apply the new variational models to explore the induced chirality of a bundle of cross-linked actin filaments.  The intuition underlying this study is based on actin stress fibers, which are bundles comprising $\sim 7 - 20$ filaments under significant tensile stress that transmit cell-wide forces during processes like cell migration \cite{fletcher2010cell,chandrasekaran2019remarkable}.  The chiral coupling between axial stretching and filament twisting, captured by the $C_{3,3}$ parameter, opens the possibility that the filaments in a stress fiber also experience significant torsion under axial stress.  We hypothesize that, due to the finite width of the actin filaments, this torsion will move the attached point of the bound cross-linker protein which will in turn pull on the other filament to which it's bound.  This will cause the peripheral filaments to tilt with respect to the central filament, such that the entire bundle structure acquires a helical pitch due to the applied tension, the twist-stretch coupling, and the bound cross-linkers.  

\begin{figure}[H]
	\begin{center}
		\includegraphics[width=10cm]{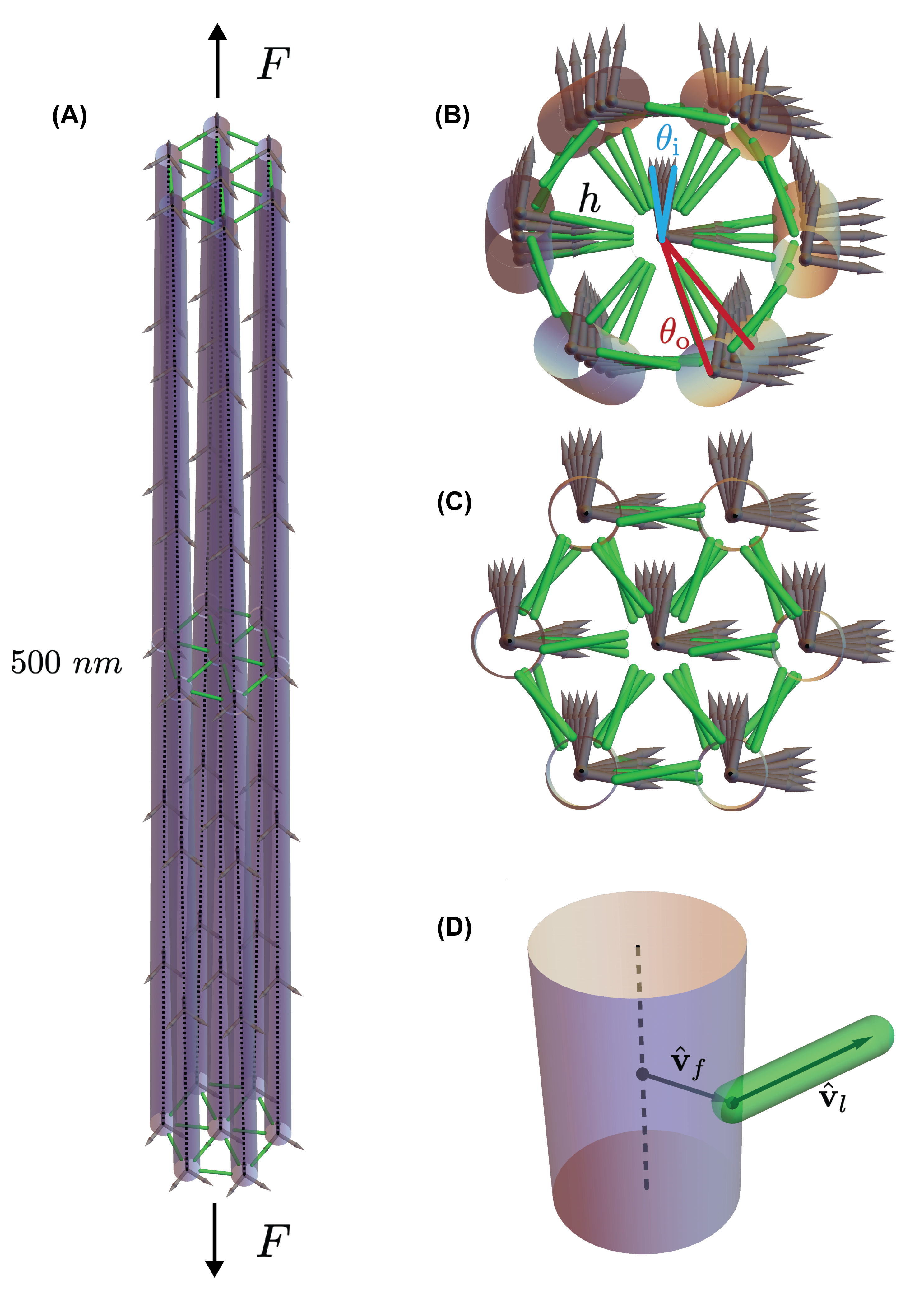}
		\caption{The structural details of the chiral bundle simulation are shown. (A)  A side-on view  is shown of a bundle of $7$ filaments.  Each filament is $500$ $\mbox{nm}$ long, lies along the $z$-axis, and has $3$ sets of cross-linkers (shown in green) attached at its ends and midpoints to each of its neighboring filaments.  A constant force $F$ is applied to each filament at both ends, causing each filament to undergo a twist due to the twist-stretch coupling when $C_{3,3} \neq 0\ \mbox{pN} \ \mbox{nm}$.  (B)  A top-down view of the same bundle is shown, illustrating more clearly the rotation of the outer filaments around the central filament. The resting length of the cross-linkers is $h$.  Red and blue wedges are drawn to illustrate the outer and inner angles $\theta_\text{o}$ and $\theta_\text{i}$ describing the bundle's twist.  It is shown how the the steric penalty ($ \varepsilon_\text{steric} \neq 0$) can cause $|\theta_\text{i}| \approx |\theta_\text{o}|$.  (C)  Without the steric penalty, the filaments tend to move inward rather than tilt and bend around the central filament, such that $|\theta_\text{i}| > |\theta_\text{o}| \approx 0$.  (D)  An illustration is shown of the unit vectors $\hat{\bm{v}}_f$ and $\hat{\bm{v}}_l$ used in the definition of the steric interaction energy, equation $(\ref{stericU})$.}
		\label{BundleCompositeStruct}
	\end{center}
\end{figure}

To explore this possible effect, we simulated $7$ filaments, each $500 \ \mbox{nm}$ long and $7$ $\mbox{nm}$ in diameter, in a bundle connected by $3$ sets of cross-linkers attached in a spoke and rim pattern, as visualized in Figures \ref{BundleCompositeStruct}.A, B, and C.  Each filament represented using the $\text{B}, \ m=1$ model with $3$ knot points.  Despite its less accurate performance under large applied loads (as shown in Figure~\ref{CompositeComparison}), this model was observed to be consistent with other models tested for this study, and we use it here because it produced the cleanest trends due to its easily minimized energy function.  The axial tension of a stress fiber was modeled by applying a constant $z$-direction force $F$ in opposite directions to both ends of every filament in the bundle.  The attached cross-linkers have a stretching energy given by
\begin{equation}
	E_\text{linker} = \frac{k_\text{linker}}{2}\left(l - h\right)^2,
	\label{linkerU}
\end{equation}
where $l$ is the cross-linker's instantaneous length and $h$ is its rest length.  With only this energy included, the peripheral filaments in the bundle tend to twist under tension and move inward toward the bundle center, allowing the lengths of the cross-linkers to achieve their rest lengths without causing the filaments to tilt (see Figure~\ref{BundleCompositeStruct}.C).  Certain actin binding proteins such as Arp2/3 are known to form relatively rigid angles with respect to the actin filament \cite{goley2006arp2}.  To account for this possibility, we also include a steric interaction which penalizes cross-linker orientations deviating from the local surface normal of the filament, with an energy given by 
\begin{equation}
	E_\text{steric} = \varepsilon_\text{steric}\left(1 - \hat{\bm{v}}_f \cdot \hat{\bm{v}}_l  \right)
	\label{stericU}
\end{equation}
where the unit vector $\hat{\bm{v}}_f$ denotes the local surface normal and $\hat{\bm{v}}_l$ denotes the unit vector pointing along the cross-linker's length from the local attachment point (see Figure~\ref{BundleCompositeStruct}.D. for an illustration of these vectors).  This steric geometric penalty has the effect that the bundle is less compressible, such that filaments are less able to move toward the bundle center and will instead tend to tilt and rotate to satisfy the linker length penalty.  The adjustable parameters of this set-up are $F$, $C_{3,3}$, $\varepsilon_\text{steric}$, $k_\text{linker}$, and $h$.  We independently varied these parameters one at a time, holding the other parameters at their default values of $F = 100 \ \mbox{pN}$, $C_{3,3} = 10^5 \ \mbox{pN} \ \mbox{nm}$, $\varepsilon_\text{steric} = 500 \ \mbox{pN} \ \mbox{nm}$, $k_\text{linker} = 10 \ \mbox{pN}/\mbox{nm}$, and $h = 8 \ \mbox{nm}$.    

We distinguish between the inner and the outer rotation of the bundle under tension.  The inner rotation $\theta_\text{i}$ is the angle through with the central filament is twisted from its minus end to its plus end, and we define $\gamma_\text{i} = |\theta_\text{i}| / \hat{L}$ as the rotation per unit length.  To measure the tilting of the peripheral filaments around the central filament, we define the outer rotation $\theta_\text{o}$ as a function of the distance $\delta r$ by which a given outer filament's endpoints are separated from each other when projected to the $xy$-plane.  If the distance from a peripheral filament's endpoint to the central filament's endpoint (i.e. the radius of the bundle) is $a$, then the outer rotation angle is defined as $\theta_\text{o} = \arccos\left(1- (\delta r)^2 /2 a^2 \right)$, given per unit length as $\gamma_\text{o} = |\theta_\text{o}| / \hat{L}$ (see Figure~\ref{BundleCompositeStruct}.B for an illustration of these angles).  The inner rotation $|\theta_\text{i}|$ will always be greater than or equal to $|\theta_\text{o}|$, and if the cross-linkers perfectly transmit the rotation of the filaments into the tilting of the outer ring, then $|\theta_\text{i}| = |\theta_\text{o}|$.

\begin{figure}[H]
	\begin{center}
		\includegraphics[width=15.6 cm]{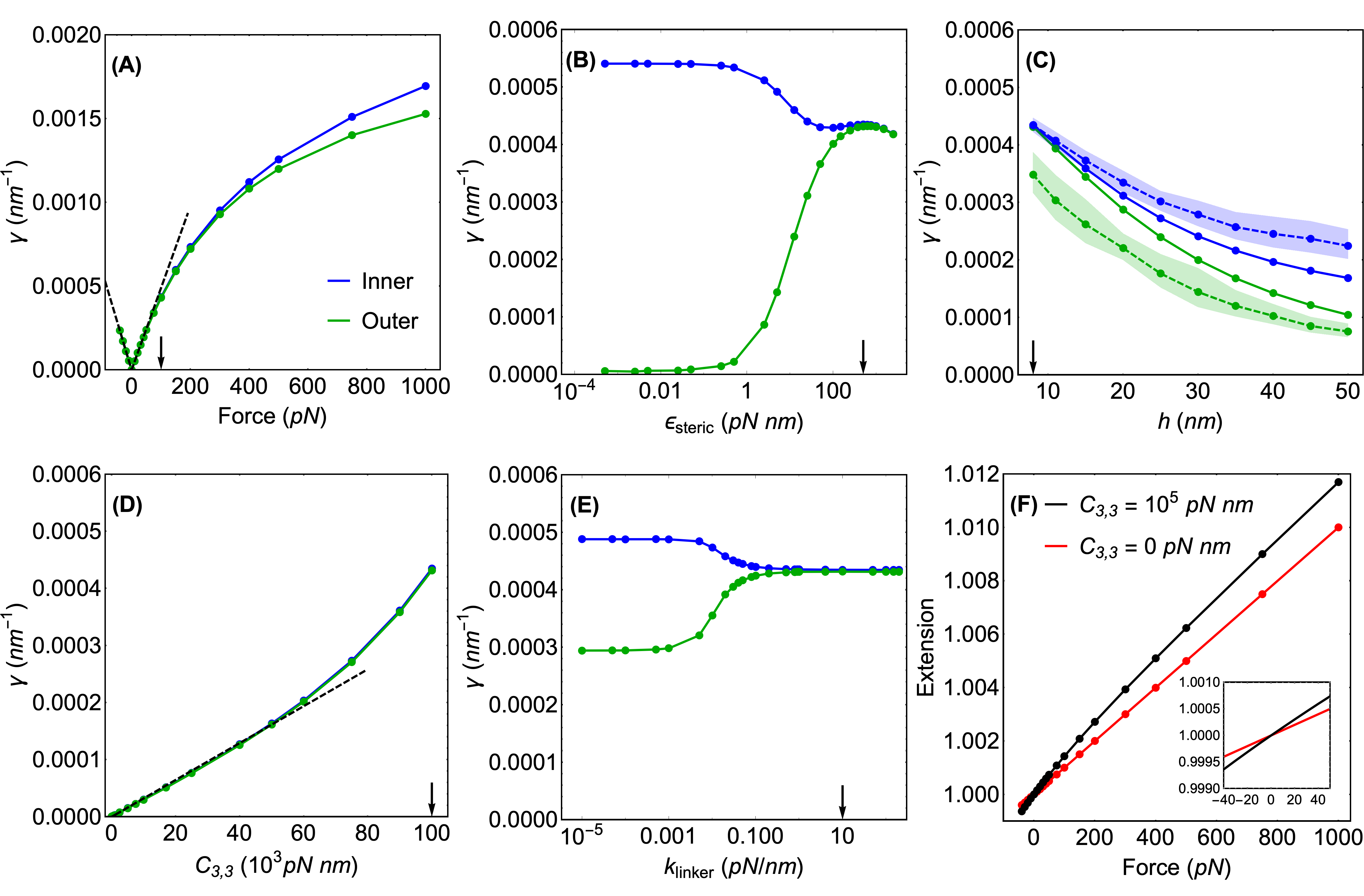}
		\caption{Results from the chiral bundle simulation are shown. 
			(A) A plot is shown of the inner (blue) and outer (green) rotation magnitude per unit length, measured in radians per $\mbox{{\rm nm}}$, as the force $F$ is varied.  The black arrow indicates the value of $F$ used when the other parameters are varied, and similarly for the arrows in the other plots. The black dashed lines show the linear response around $F=0 \ \mbox{{\rm pN}}$.  (B)  A plot is shown of the inner and outer $\gamma$ as $\varepsilon_\text{steric}$ is varied.  (C)  A plot is shown of the inner and outer $\gamma$ as $h$ is varied. The dashed line and shaded area indicate the mean and standard deviation of $100$ realizations of randomly placing $3$ cross-liners between the pairs of filaments, rather than attaching them at the ends and filament midpoint.  (D)  A plot is shown of the inner and outer $\gamma$ as $C_{3,3}$ is varied.  (E)  A plot is shown of the inner and outer $\gamma$ as $k_\text{linker}$ is varied.  (F)  A plot is shown of the relative extension $L / \hat{L} $ as $F$ is varied for two values of the coupling parameter $C_{3,3}$.  The inset is a blow-up around $F=0\ \mbox{{\rm pN}}$ showing the crossover behavior.}
		\label{BundlePlotsMPF}
	\end{center}
\end{figure}

Several notable trends are observed in this study, displayed in Figure~\ref{BundlePlotsMPF}.  First, we find a transition from a linear dependence of both $\gamma_\text{o}$ and $\gamma_\text{i}$ on $F$ to sublinear dependence, at which point the outer and inner rotations also begin to separate from each other so that $\gamma_\text{i} > \gamma_\text{o}$ (Figure~\ref{BundlePlotsMPF}.A). This behavior is also symmetric about $F=0$, with compression inducing twisting in an approximately equal but opposite amount to stretching for small loads.  These observations qualitatively agree with the intuition of linear response for small disturbances transitioning to nonlinear response for large disturbances.  We also find a transition from linear to superlinear dependence on $C_{3,3}$ (Figure~\ref{BundlePlotsMPF}.D).  

Second, we find a strong nonlinear dependence of $\gamma_\text{o}$ and $\gamma_\text{i}$ on $\varepsilon_\text{steric}$ resembling a second-order phase transition (Figure~\ref{BundlePlotsMPF}.B) \cite{binney1992theory}.  Below an apparent threshold around $\varepsilon_\text{steric} \approx 0.1 \ \mbox{pN} \ \mbox{nm}$ the applied force rotates the central filament by a fixed amount and the peripheral filaments rotate and move inward to satisfy the cross-linker length energy penalty. Above this threshold, and in a continuous manner, the steric penalty causes the filaments to tilt and bend rather than move inward to satisfy the cross-linker length penalty, causing $\gamma_\text{o}$ to increase and $\gamma_\text{i}$ to decrease.  Above an upper threshold around $\varepsilon_\text{steric} \approx 100 \ \mbox{pN} \ \mbox{nm}$ this trend saturates, and only small, though interestingly non-monotonic, changes are observed in $\gamma_\text{i}$ and $\gamma_\text{o}$ which now roughly coincide.  Similar behavior is found for the dependence on $k_\text{linker}$, although the outer rotation below the lower transition threshold $\sim 0.01 \ \mbox{pN}/\mbox{nm}$ is constant at a finite value, not zero, implying that the steric penalty alone can cause outer rotation of the bundle (Figure~\ref{BundlePlotsMPF}.E).  
\renewcommand{\floatpagefraction}{.8}

Third, we find that as the cross-linker resting length $h$ is increased, both $\gamma_\text{i}$ and $\gamma_\text{o}$ monotonically decrease, and $\gamma_\text{i}$ grows relative to $\gamma_\text{o}$ suggesting less effective transduction of inner rotation to outer rotation for large linker lengths (Figure~\ref{BundlePlotsMPF}.C).  We also tested the effect of randomly placing the cross-linkers between the filaments rather than at the ends and midpoint of the filaments, controlling for the number of cross-linkers between each pair.  For each value of $h$, we sampled 100 realizations of cross-linker positions with uniform probability along the filament lengths.  We expected that the ordered (but statistically unlikely) arrangement of cross-linkers enhances the transduction of inner to outer rotation because the forces throughout the bundle are highly coordinated.  The rare configuration with ordered cross-linkers is indeed more effective at causing outer rotation than the typical random configuration, as shown by the dotted lines in Figure~\ref{BundlePlotsMPF}.C which have larger $\gamma_\text{i}$ and smaller $\gamma_\text{o}$ for all $h$.  

Finally, we tested how the mechanical coupling between twisting and stretching affected the force-extension curve of the bundle, shown in Figure~\ref{BundlePlotsMPF}.F.  The equilibrated length $L$ of the central filament was measured as a function of the pulling force and divided by its initial value $\hat{L} = 500 \ \mbox{nm}$ to give the relative extension.  We found that a non-zero $C_{3,3}$ allows for greater extension and greater compression for a given force $F$.  For $C_{3,3} = 0\ \mbox{pN} \ \mbox{nm}$ the force-extension curve is perfectly linear, while for $C_{3,3} \neq 0 \ \mbox{pN} \ \mbox{nm}$ it smoothly interpolates between an asymptotically linear regime for $F \gg 0 \ \mbox{pN}$ and a nonlinear crossing regime around the point $C_{3,3} = 0\ \mbox{pN} \ \mbox{nm}, \ F=0 \ \mbox{pN}$.  

\subsection{MEDYAN Implementation}

	As a final application, we implemented the GC model into MEDYAN \cite{popov2016medyan}.  This implementation consists of several new modelling choices, which we describe in detail in the Supplementary material.  These new modelling choices have to do with allowing for chemical reactions, such as cross-linkers and molecular motors binding to filaments and filament polymerization and depolymerization reactions, using the new filament mechanical model presented in this paper.  These additional chemical considerations, which allow the current mechanical model to be incorporated in a versatile active matter simulation platform, should significantly expand the model's usefulness in studying cytoskeletal dynamics.

\begin{figure}[H]
	\begin{center}
		\includegraphics[width=\textwidth]{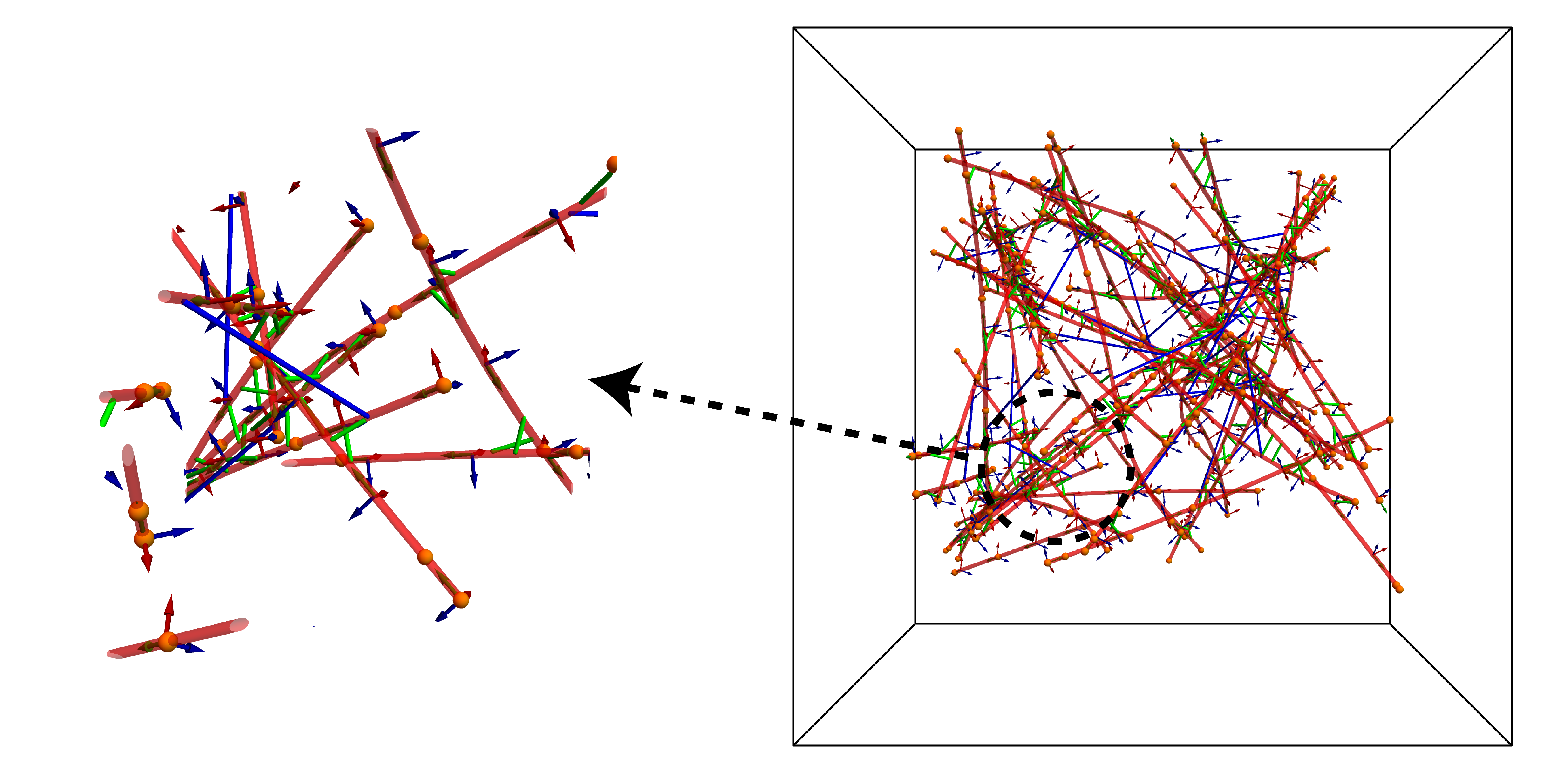}
		\caption{A visualization of a contractile actomyosin network from a MEDYAN simulation is shown, in which actin filaments are drawn as red tubes, myosin minifilaments are shown as blue cylinders, and $\alpha$-actinin cross-linkers are shown as green cylinders.  The left panel shows a blow-up of a region of the full network, which is shown in the right panel, in which additional detail of the surface-bound minifilaments and cross-linkers can be seen.  The director triads, which are defined at every point on the actin filaments, are visualized as periodically spaced red, green, and blue arrows.  The orange spheres represent the knot coordinate points, and the black box represents the simulation boundary.  The snapshot is taken after $150\;$s of simulated time.}
		\label{NetworkComposite}
	\end{center}
\end{figure}

	In Figure~\ref{NetworkComposite} we show a snapshot from a MEDYAN simulation, in which an actomyosin network comprising physiological concentrations of actin, myosin (non-muscle myosin IIA) and cross-linkers ($\alpha$-actinin) has undergone a network-wide contraction away from the simulation boundaries.   This motor-driven contraction is in keeping with well-documented behavior of actomyosin networks at these concentrations \cite{popov2016medyan, floyd2019quantifying, linsmeier2016disordered}.  The key point is that the network in Figure~\ref{NetworkComposite} has binding molecules attached to the surfaces of the actin filaments rather than their backbones, allowing for network-level shearing and twisting forces, filament rotational dynamics, and chiral phenomenon to be studied \textit{in silico}.  We report here only the feasibility of implementing the GC model into a network-level simulation platform like MEDYAN, rather than any trends observed using this implementation which we plan to explore in depth in future works.

\section{Discussion}\label{Discussion}
An important physical feature currently missing from large-scale mechanochemical simulations of cytoskeletal networks is the finite width of the filaments.  This precludes studying effects in which filaments can rotate or shear in response to forces in the network, arising for instance from bound molecular motors or polymerization against a boundary.  To address this, we have presented in this paper a set of options for parameterizing the configurations of filaments with finite widths in a computationally efficient way, requiring only a small number of free model parameters.  We focused here on variational models, in which we explore functional forms for a filament's mechanical degrees of freedom in order to efficiently find their equilibrated configurations.  We introduced two main classes of functional forms: one in which a sequence of splines is used to parameterize the filament configuration, and one in which a sequence of geodesic curves in the space of orthogonal rotation tensors is used.  In the spline-based approach, the intractable expressions for the strain components necessitated a small-angle expansion of the energy function.  This was not necessary in the geodesic approach due to simplifications in the expressions of the strain components arising from the geodesic curve parameterization.  This small-angle approximation is commonly made in treatments of elastic rods, but is shown here to give rise to significant deviations from expected behavior under large applied loads \cite{de2009cofilin,o2017modeling}.  By avoiding the small-angle approximation, the geodesic approach showed close agreement even under large applied loads when compared with accurate yet computationally expensive dynamical simulations, whereas the spline-based approach exhibited systematically smaller deformations compared to the dynamical solution.  This systematic error may also be partly explained by examining the $\hat{s}$-profiles of the strain components in the various models, as discussed in the Supplementary material, Section \ref{Strain profiles}.  

One practical concern in implementing the various functional forms introduced here is the associated computational cost of evaluating the filament position and energy and of numerically minimizing the energy.  A related issue of these models is their locality, or the dependence of local quantities on either only ``nearby'' parameters of the model rather than on almost all parameters of the model.  For instance, in the spline-based models smoothness is enforced by relating the control points of later segments to those of earlier segments.  This causes the expression for the position on segment $i$ to depend on all parameters up to that segment, so that the complexity of the model grows faster than linearly with the number of knot points.  It should be possible to use B-splines to alleviate this non-locality in future developments \cite{shikin1995handbook,prautzsch2002bezier}.  Non-locality is more inherent in the geodesic model, arising from the expression of the backbone curve $\bm{r}(\hat{s})$ as an integral up to $\hat{s}$ of the tangent $\mbox{d}\bm{r}/\mbox{d}\hat{s}$.  While closed form solutions for this integral are straightforwardly obtained (see the Supplementary material), there is no way to avoid this dependence of the position on segment $i$ on the parameters up to segment $i$.  Measurements of the time taken to evaluate the energy function as $N_\text{k}$ is varied for each model nonetheless show that the geodesic models may be more computationally efficient than the spline-based model, as discussed in the Supplementary material, Section \ref{Timing and accuracy of the models}.  

Various extensions to the models presented here are possible.  For the spline-based models, it was mentioned that B-splines may be used to improve locality, and exponential splines could also be used to increase expressivity by allowing for both polynomial and exponential contributions to the filament functional form \cite{shikin1995handbook,prautzsch2002bezier,spath1969exponential}.  One could also mix the geodesic and spline-based approaches.  For instance, in the geodesic models one can relax the constraint that the components $\zeta_\alpha(\hat{s})$ of $\mbox{d}\bm{r}/\mbox{d}\hat{s}$ are constant on the segments; $\zeta_\alpha(\hat{s})$ could instead be a spline function in $\hat{s}$ on the segment and the energy terms involving $\zeta_\alpha(\hat{s})$ could still be found exactly (see the Supplementary material, Section \ref{Functions in the goedesic models}).  Other functional forms not considered here could also be investigated.  Rather than using splines to parameterize the Euler angles of $\bm{Q}$, splines could be used to parameterize curves of quaternions or other representations for $\bm{Q}$ \cite{altmann2005rotations}.  In principle one could also allow the knot coordinates $\hat{s}_i$ to become free model parameters, so that the segment lengths are adjustable during minimization.  Additionally, one may use the functional forms presented here but adopt a dynamical, rather than variational, approach to study filament mechanics.  Considering the free model parameters of these functional forms to be generalized mechanical coordinates, one could derive equations of motion giving the time evolution of the filament's configuration using Hamiltonian or Langevin dynamics \cite{goldstein2002classical,risken1996fokker}.  This can offer a way to endow a filament with all mechanical degrees of freedom of the Cosserat model in time integration-based simulations of semi-flexible polymer networks, while preserving the computational efficiency of tracking only a handful of free model parameters \cite{nedelec2007collective,freedman2017versatile,kim2009computational}.  Finally, our work is based on the Cosserat theory of elastic rods which is more general than the Kirchoff theory, but less general than the theory of Green and Naghdi which allows in-plane shearing of the rod's cross-sections \cite{green1966general,green1995unified}.  Accommodating in-plane shearing deformations considerably complicates the mathematics by introducing non-orthogonal local directors, and we expect that it contributes only minor corrections to the dynamics of filaments like actin.  However, future work may apply this more general approach to study biopolymer mechanics.

In this paper, we have considered a coarse-grained representation of an actin filament which has a constant circular cross-section and lacks monomer-level resolution.  In the accompanying paper, Ref.~\citenum{gunaratne2022stretching}, we describe a finger-grained monomer-level model of an actin filament that preserves the helical filament microstructure, and we develop a method for smoothly connecting the monomeric model to the constant cross-section model presented here.  This multi-resolution modeling approach allows for fine control over the trade-off between biological detail and computational expense.  In Ref. \citenum{gunaratne2022stretching} we also discuss in detail issues of parameterization, which was treated only briefly here (see Supplementary material), as well as validations of these models using direct comparison to experimental measurements of actin filament configurations.  Therefore, while some important chemical detail has been omitted in the present paper, we discuss in Ref. \citenum{gunaratne2022stretching} how this detail can be built back into the model in a systematic manner.

An exciting future application of efficient computational models of finite-width filaments will be to investigate emergent chiral symmetry breaking in active, self-organizing cytoskeletal networks.  Our simulation of a chiral filament bundle can be viewed as a preliminary investigation into this topic, showing that chirality in the mechanical compliance of individual finite-width filaments (as encoded in the parameter $C_{3,3}$) can give rise through surface-bound cross-linkers to chiral rotation of a multi-filament bundle.  Other mechanisms by which broken chiral symmetry can propagate to larger spatial scales may be studied in more complete simulations of motorized cytoskeletal networks, for instance using a future version of MEDYAN augmented to use a Cosserat model for filaments \cite{popov2016medyan}.   In addition, such network-level simulations could explore the effect of cofilin on cytoskeltal dynamics.  It has been shown that cofilin molecules bind cooperatively to actin filaments and induce a torsional strain that leads to filament severing \cite{de2009cofilin,mccullough2011cofilin}.  This non-trivial mechanical effect could be realistically accounted for in simulation using the finite-width models presented here.

\clearpage
\appendix

\section{Supplementary results}\label{Supplementary results}
\subsection{Euler buckling study}\label{Euler buckling study}
\begin{figure}[H]
	\begin{center}
		\includegraphics[width=8 cm]{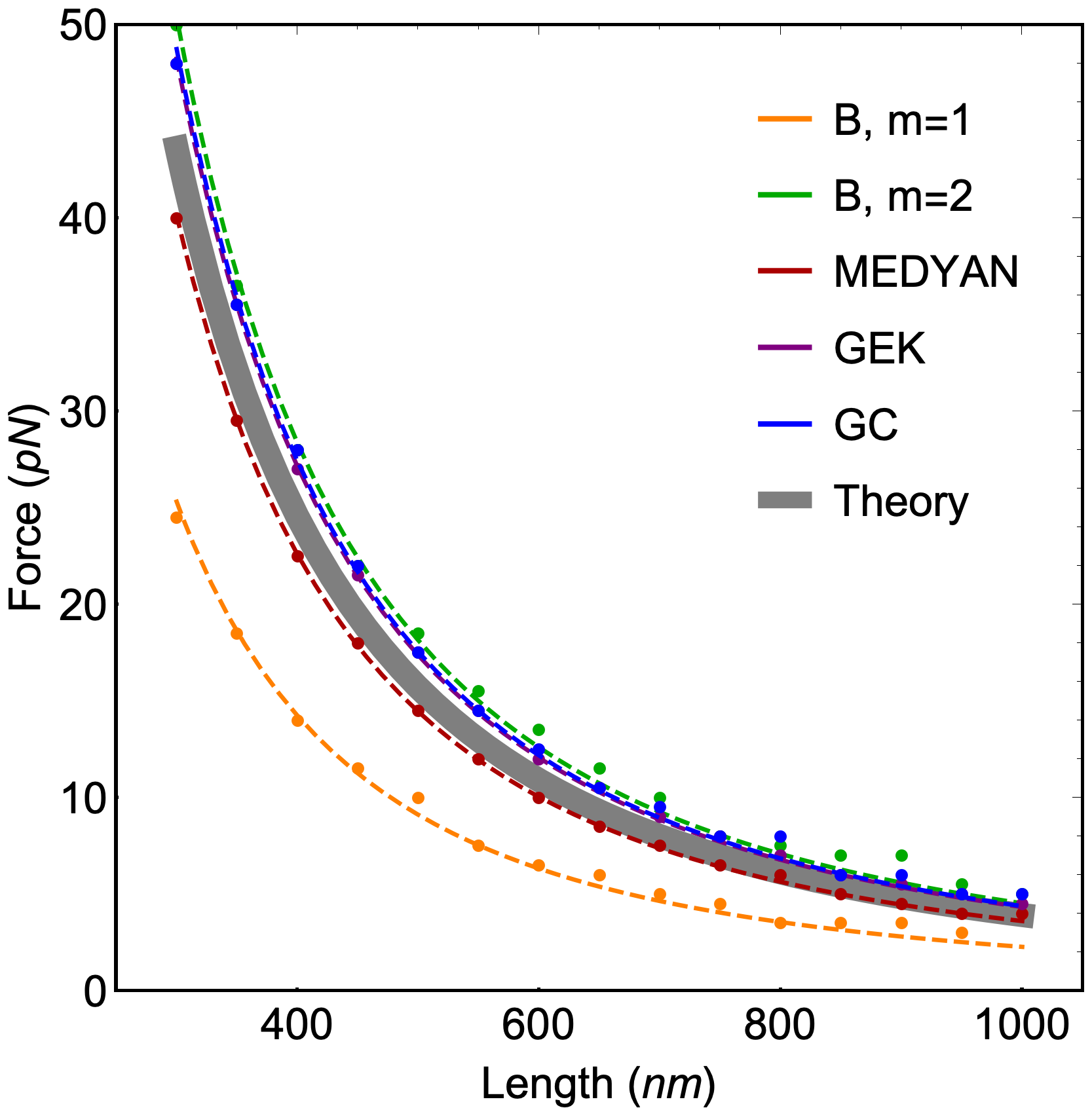}
		\caption{The Euler critical buckling force $F_\text{c}$ is shown for various filament lengths $\hat{L}$ for five variational models.  The scatter plot data indicates the minimal value of $F$ producing a buckled configuration for that value of $\hat{L}$, and through these points the curve in Equation (\ref{Eulerbucklingfit}) is fit to determine the value of $a$ for that model.   The model labels are described in the main text.  The gray curve corresponds to the value $a=1$.}
		\label{EulerBucklingFig}
	\end{center}
\end{figure}

As a test of the variational models introduced in this paper, we computed the force needed to buckle the filaments as a function of their length.  For inextensible and unshearable elastic rods, there is a formula by Euler for the critical buckling force (i.e. the minimal force causing the filament to buckle):
\begin{equation}
	F_\text{c} = \frac{\pi^2 B_{1,1}}{\left(K_{E} \hat{L}\right)^2},
	\label{Eulerbuckling}
\end{equation}
where $K_{E}$ is a numerical constant depending on the constraints applied at the filament endpoints, $\hat{L}$ is the material length of the filament, and $B_{1,1}$ is the bending stiffness appearing in Equation \ref{epsilondef} of the main text \cite{howard2001mechanics}.  In this study, the filaments lie initially on the $z$-axis and the minus-ends are constrained to the origin but can freely rotate.  A constant (gravitational) force $F$ is applied to the plus-end of the filament in the $-z$ direction and the $x y$ coordinates of the plus-end are constrained to lie above the origin, but the filament may rotate at that end.  For this set-up, $K_\text{E} = 1$.  All mechanical parameters ($\bm{S}$, $\bm{B}$, and $\bm{C}$) were chosen to correspond to actin (see the Parameterization section of the Supplementary Material), and the inextensibility and unshearability condition was imposed by setting $S_{3,3}$ and $S_{1,1}$ to $10^3$ times their usual values. We also set $\bm{C} = \bm{0}$.  We tested $14$ values of $\hat{L}$, from $300 \ \mbox{nm}$ to $1 \ \mu$m in increments of $50 \ \mbox{nm}$, and for each $\hat{L}$ we tested $100$ values of $F$, from $0.5 \ \mbox{pN}$ to $50 \ \mbox{pN}$ in increments of $0.5 \ \mbox{pN}$.  The initial coordinates of the filament were given a small random perturbation around the initially straight configuration to break the initial symmetry and allow buckling to occur.  A filament was judged to be buckled if its midpoint displacement or energy exceeded certain threshold values, which were not found to be very sensitive parameters.  In Figure~\ref{EulerBucklingFig} we plot the minimal values of $F$ for each $\hat{L}$ which produced a buckled filament.  Through these points we fit curves of the form 
\begin{equation}
	F_\text{c,fit} = a \frac{\pi^2 B_{1,1}}{\hat{L}^2},
	\label{Eulerbucklingfit}
\end{equation}
for the prefactor $a$.

All models obey the $F_\text{c} \propto \hat{L}^{-2}$ scaling predicted by the Euler theory, but the quantitative accuracy, judged by the deviation of $a$ from 1, varies across the models.  The value of $\lvert a-1\rvert$ determined for each model is summarized in Table \ref{TableBuckling}.  Interestingly, the original zero-width MEDYAN model has the best agreement with the Euler buckling theory, although the geodesic models and the second order spline-based model also agree well.  The first order spline-based model significantly underestimates the buckling force, however, indicating the insufficiency of the small-angle expansion for large filament deformations.  Finally, we note that the dynamical model also used in this paper has already been shown to exhibit excellent agreement with the theoretical buckling prediction and is not tested here \cite{gazzola2018forward}.

\begin{table}[H]
	\begin{center}
		\begin{tabular}{|c c|} 
			\hline
			Model  & $\lvert a-1\rvert$  \\ 
			\hline\hline
			B, $\ m=1$ & 0.420 \\
			B, $\ m=2$ & 0.155 \\
			GEK & 0.106 \\
			GC & 0.117 \\
			MEDYAN & 0.080 \\
			\hline
		\end{tabular} 
		\caption{The deviation of the fitted prefactor $a$ from $1$ is shown for the five variational models tested.}
		\label{TableBuckling}
	\end{center}
\end{table}

\subsection{Description of test cases}\label{Description of test cases}
We tested three cases of filament lengths and attached loads.  In the test case (A) (Figure~\ref{CompositeComparison}.A), the length of the filament is $500 \ \mbox{nm}$ and it lies along the $z$-axis.  Four cross-linkers, modeled as harmonic springs, are attached to the filament backbone at $\hat{s}^\text{b} = 0, \ 167, \ 375$, and $500 \ \mbox{nm}$; the other endpoints have $x$ coordinates at $x = 100, \ -100, \ 100$, and $-100 \ \mbox{nm}$, respectively.  In the test case (B) (Figures \ref{CompositeComparison}.B and \ref{CompositeComparison}.C), the length of the filament is $100 \ \mbox{nm}$ and it lies again along the $z$-axis.  Three cross-linkers are attached to the backbone at $\hat{s}^\text{b} = 0, \ 33$, and $100 \ \mbox{nm}$; the other endpoints are at $x = 30, \ -30$, and $30 \ \mbox{nm}$, respectively.  Test case (C) is identical to the test case (B), except the other cross-linker endpoints are at $x = 100, \ -100$, and $100 \ \mbox{nm}$, respectively.  Each cross-linker has an equilibrium length of $8 \ \mbox{nm}$ and a spring constant of $10 \ \mbox{pN} / \mbox{nm}$.  For the five variational models, we used $N_\text{k} = 5$ for each test case, and for the dynamical model we used for each test case a segment length of $0.1 \ \mbox{nm}$, a time-step of $10^{-4} \ \mbox{s}$, and a total simulation time of $50 \ \mbox{s}$.  The parameters in the matrices $\bm{S}$ and $\bm{B}$ were chosen to describe actin filaments, as described in the Supplementary Material.  For these test cases, the coupling matrix $\bm{C}$ was set to zero.  The variational models were implemented in Mathematica, and minimization of $E_\text{tot}(\mathcal{K})$ was done using a Mathematica library implementation of the conjugate gradient algorithm \cite{mathematica,nocedal2006numerical}. For all comparisons to the variational models, we used a dissipation constant of $\gamma_\text{d} = 10 \ \mbox{pN} \ \mbox{s} / \mbox{nm}$, and we checked that the dynamical solution had indeed converged and represented an equilibrated configuration.  The dynamical model of Gazzola et al. was implemented in MATLAB \cite{MATLAB:2021a}.

\subsection{Timing and accuracy of the models}\label{Timing and accuracy of the models}
\begin{figure}[H]
	\begin{center}
		\includegraphics[width=8 cm]{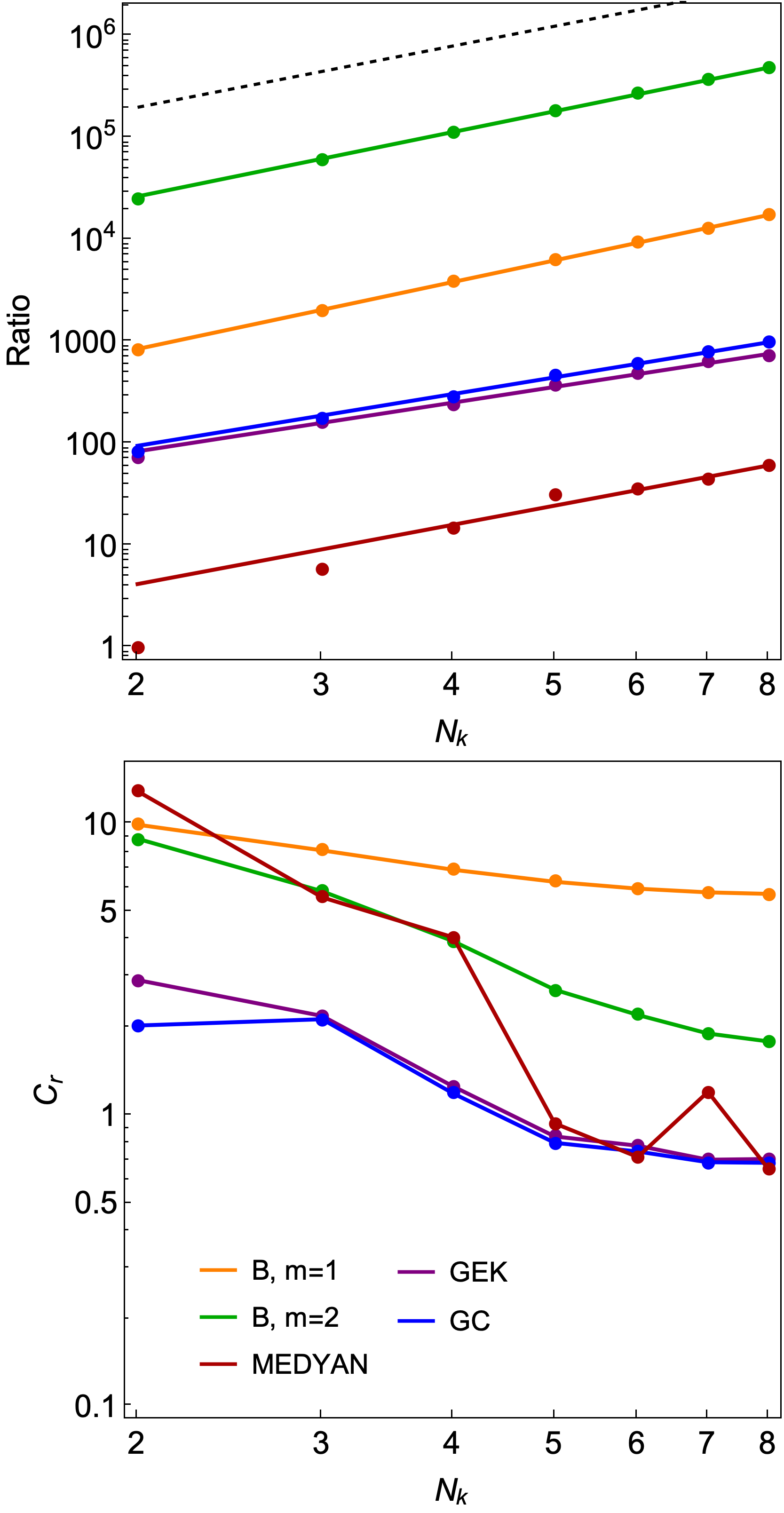}
		\caption{The timing and accuracy of the variational models is displayed as the number of knot points $N_\text{k}$ is varied.  The top panel shows as scatter plot data the mean time taken to evaluate the energy function $E(\mathcal{K})$ for each condition, shown as a ratio over the time taken to evaluate the MEDYAN energy function for $N_\text{k} = 2$.  For each model, the scatter plot data is fit by a function of the form $a N_\text{k}^b$.  The data is shown on a log-log plot, and the black dashed line has a slope of $2$.  The bottom panel shows as joined scatter plot data the accuracy of each model as a $N_\text{k}$ is varied for the third test case in Figure~\ref{CompositeComparison} of the main text.  }
		\label{Timing}
	\end{center}
\end{figure}

Here we study how varying the number of knot points $N_\text{k}$ affects the accuracy and computational timing of the variational models.  We measured the CPU time taken to evaluate the energy function of the filament $E(\mathcal{K})$ for each of these models and choices of $N_\text{k}$.  All implementations are done in Mathematica \cite{mathematica}.   Rather than report the absolute timing of these function evaluations, we report the timing relative to the fastest time obtained (for the $N_\text{k} = 2$ original MEDYAN model).  Each timing data point is an average over $30$ samples.  We also tested for each choice of $N_\text{k}$ the model accuracy for the test case (C) in Figure~\ref{CompositeComparison} of the main text.  We measured this accuracy using the RMSD backbone distance metric $ C_r\left(\bm{r}^A(\hat{s}), \bm{r}^B(\hat{s})\right)$ defined in Equation \ref{Crdef} of the main text, where $\bm{r}^B(\hat{s})$ for all comparisons is the finely-discretized dynamical model solution.  The results are displayed in Figure~\ref{Timing}.  We note that the CPU time needed to stably propagate the dynamical model for long enough to achieve an equilibrated filament configuration is orders of magnitude larger than the time needed to numerically minimize any of the variational models, highlighting the extreme gain in computational efficiency from using the variational approach.

\begin{table}[H]
	\begin{center}
		\begin{tabular}{|c c|} 
			\hline
			Model  & $b$  \\ 
			\hline\hline
			B, $\ m=1$ & 2.18 \\
			B, $\ m=2$ & 2.10 \\
			GEK & 1.59 \\
			GC & 1.69 \\
			MEDYAN & 1.93   \\
			\hline
		\end{tabular} 
		\caption{The scaling exponent of CPU time with $N_\text{k}$ for the five variational models tested here.}
		\label{TableScaling}
	\end{center}
\end{table}

We see that the function evaluation timing for all models exhibit approximately a $N_\text{k}^b$ scaling, with $b \approx 2$ as shown in Table \ref{TableScaling}.  We emphasize that the timing is studied only for the function evaluation, rather than for the actual minimization of the energy which may depend sensitively on the minimization algorithm employed.  The accuracy increases monotonically for each model as $N_\text{k}$ is varied (with a small exception for the MEDYAN model at large vales of $N_\text{k})$.  In addition, it is found that each model's accuracy tends to plateau after $N_\text{k}$ is made sufficiently large.  However, the accuracies at which the different models plateau varies significantly.  While the MEDYAN model is observed to obtain realistic configurations for large $N_\text{k}$, it is a zero-width model which does not include shearing and twisting of the filament.  Thus the geodesic models attain the best accuracy while allowing for filaments to have all mechanical degrees of freedom in the Cosserat theory.

\subsection{Strain profiles}\label{Strain profiles}
\begin{figure}[H]
	\begin{center}
		\includegraphics[width=16 cm]{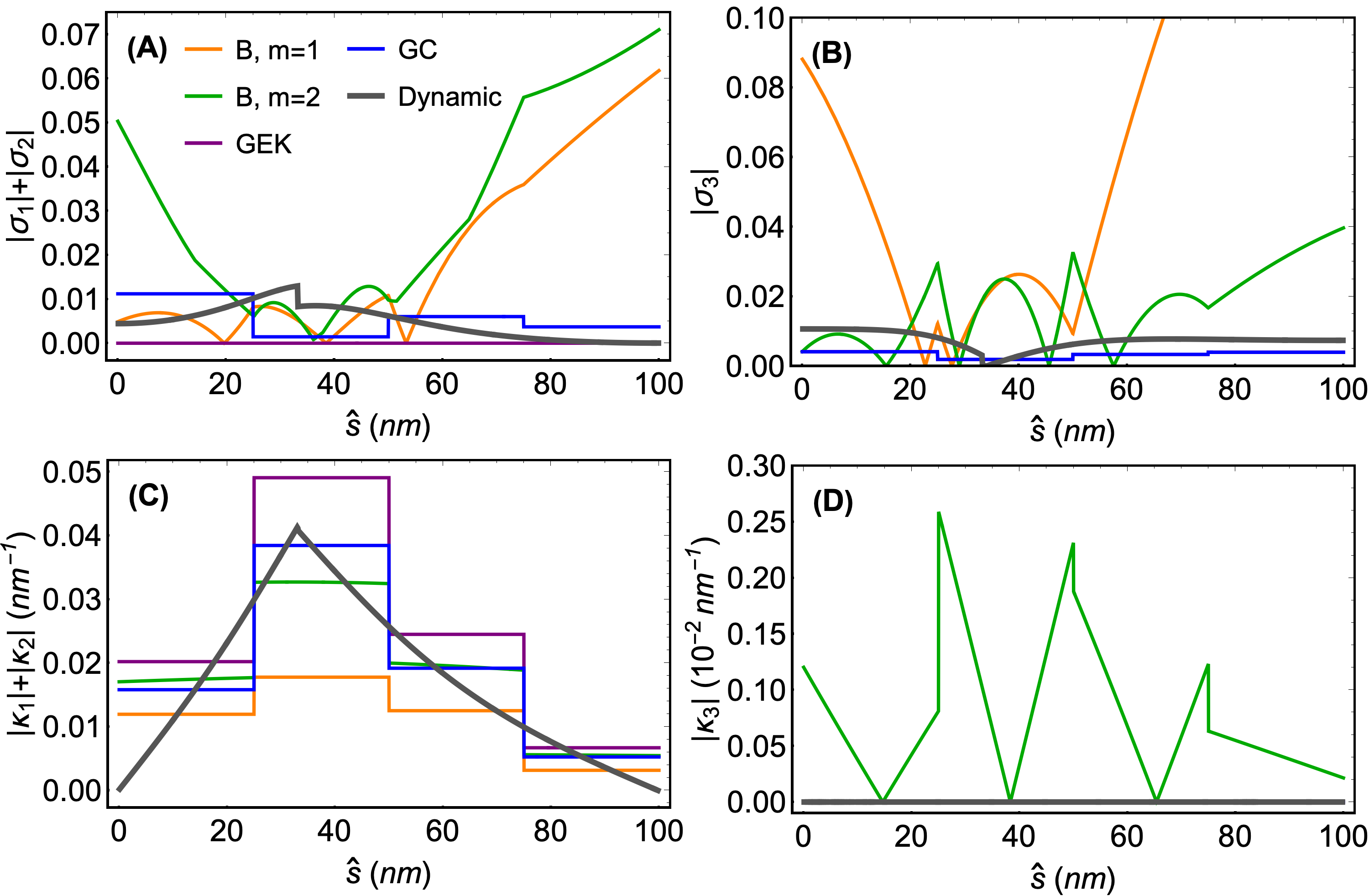}
		\caption{A comparison is shown of strain profiles along an equilibrated filament in different models for the third test case used in Figure~\ref{CompositeComparison} of the main text. (A) The total absolute shearing strain is shown as a function of the reference arc-length $\hat{s}$. (B) The absolute stretching strain is shown as a function of $\hat{s}$. (C) The total absolute bending strain is shown as a function of $\hat{s}$. (D) The absolute twisting strain is shown as a function of $\hat{s}$.}
		\label{StrainProfile}
	\end{center}
\end{figure}

To understand in greater detail how the various models differ in their representations of the filament configurations, we plotted the filament strains $\kappa_\alpha(\hat{s})$ and $\sigma_\alpha(\hat{s})$ (defined in equations (\ref{kappadef}) and (\ref{sigmadef}) of the main text) along the reference arc-length $\hat{s}$.  We used test case (C) of Figure~\ref{CompositeComparison} in the main text, including the variational models as well as the finely-discretized dynamical model.  As displayed in Figure~\ref{StrainProfile}, for the spline-based models with $m=1$ or $m=2$, the shearing and stretching strains $\sigma_\alpha$ on each segment are highly non-uniform and quite large. This, when squared and integrated, creates a large stretching and shearing energy penalty. On the other hand, the geodesic models by construction have uniform $\sigma_\alpha$ on each segment, which apparently agrees better with the true strain profile of the dynamical model.  If the shearing strain profiles that are possible to express using the spline-based model are not easily matched against those of the true filament, then the shearing energy penalty may artificially restrict the spline-based filament configurations.  This could then explain the systematically smaller filament deformations observed in Figure~\ref{CompositeComparison} of the main text.  We observe that all models have similar bending strains $\kappa_1$ and $\kappa_2$.  It is evident that the GC model has smaller bending strain than the GEK model, due to the possibility in the GC model of loading some strain into the shearing deformation in addition to the bending deformation. Interestingly, on this bent filament, the second order spline-based model produces a non-zero twisting strain $\kappa_3$.  This is likely due to the complexity of the corresponding energy function which gives rise to some artefacts during the numerical minimization procedure.  We note that this erroneous twisting strain is small compared to the bending strains, and thus represents only a slight deviation from expected behavior.

\section{Supplementary methods}\label{Supplementary methods}
\subsection{Parameterization}\label{Parameterization}

Here we describe how the diagonal elements of $\bm{B}$ and $\bm{S}$, which appear in the expression for the energy density in the main text, are determined for actin filaments.  We use the following relations, which may be found in Table 1 of Ref. \citenum{gazzola2018forward}:
\begin{eqnarray}
	B_{1,1} &=& E_\text{mod} I_{1,1} \nonumber \\
	B_{2,2} &=& E_\text{mod} I_{2,2} \nonumber \\
	B_{3,3} &=& G_\text{mod} I_{3,3} \nonumber \\
	S_{1,1} &=& \alpha_\text{c}G_\text{mod}A \nonumber \\
	S_{2,2} &=& \alpha_\text{c}G_\text{mod}A \nonumber \\
	S_{3,3} &=& E_\text{mod} A. \nonumber 
\end{eqnarray}
In these equations, $E_\text{mod}$ is Young's modulus, $G_\text{mod}$ is the shear modulus, $\alpha_\text{c}$ is a geometrical constant equal to $4/3$ for circular cross-sections, $A$ is the cross-sectional area, and $I_{\alpha,\beta}$ are elements of the second (or polar) moment of inertia tensor.  The diameter of an actin filament is in the range of 5-7 $\mbox{nm}$, placing its cross-sectional area $A$ in the range of 40-75 nm$^2$ \cite{grazi1997diameter}.  We use $A=50$ nm$^2$ throughout.  For a circular cylinder, the second moment of inertia tensor is $\bm{I} = (A^2/4 \pi) \text{diag}(1,1,2)$ \cite{ugural2003advanced}.  The Young's modulus has been estimated for actin filaments as $ E_\text{mod} \approx 2$ GPa,  however the shear modulus has not to our knowledge been directly measured \cite{kojima1994direct}.  To estimate the shear modulus of actin we use the formula $G_\text{mod} = E_\text{mod} /2(1 + \nu)$, where $\nu$ is Poisson's ratio \cite{landau1965theory}.  $\nu$ has been estimated for actin as $0.4$ \cite{tseng2002functional,kojima1994direct}.  This gives $G_\text{mod} \approx  0.7$ GPa.  

\subsection{$\mathbf{Q}(\hat{s})$ and $\mathbf{r}(\hat{s})$ in the geodesic models}\label{Functions in the goedesic models}

As described in the main text, in the geodesic models there are $N_\text{k}$ independent rotation tensors $\bm{Q}_i$ at each knot coordinate $\hat{s}_i$.  Between these knot points, the tensors are given by
\begin{equation}
	\bm{Q}_i(q) = \exp\left(q \ln \left(\bm{Q}_{i+1} \bm{Q}^T_i\right)\right) \bm{Q}_i,
	\label{eqtensorcurve}
\end{equation}
where $q(\hat{s};\hat{s}_{i},\hat{s}_{i+1}) = (\hat{s} - \hat{s}_i)/\hat{L}_i$ runs from $0$ to $1$ along the segment arc-length.  $q$ can be converted to the global arc-length coordinate $\hat{s}$ using $\hat{s} = \hat{s}_i + q \hat{L}_i$.  The global tensor curve $\bm{Q}(\hat{s})$, which is piecewise defined by the above equation, is continuous but not smooth at the knot points.  Smoothness could be enforced be requiring $\mbox{d} \bm{Q} / \mbox{d}\hat{s}$ to be continuous at the knot points, leading to equations relating the tensors of consecutive segments and reducing the number of free model parameters, but for now we do not enforce this. We use the axis-angle parameterization for each tensor $\bm{Q}_i$: for each $i$ we have $\theta^\text{Ax}_i$ and $\bm{u}^\text{Ax}_i$ so that $\bm{Q}_i = \exp\left(\text{skew} (\theta^\text{Ax}_i  \bm{u}^\text{Ax}_i)\right)$.  The axis $\bm{u}^\text{Ax}_i$ must be a unit vector, which we take to be parameterized by the polar and azimuthal angles $\beta^\text{Ax}_i$ and $\gamma^\text{Ax}_i$. Thus the collection of angles $\theta^\text{Ax}_i, \ \beta^\text{Ax}_i,$ and $\gamma^\text{Ax}_i$ parameterizes $\bm{Q}_i$ for each $i$. 

Equation (\ref{eqtensorcurve}) can be expressed using the Rodrigues formula as (dropping the $\text{Ax}$ superscript)
\begin{equation}
	\bm{Q}_i(q) = \left(\cos (q \tilde{\theta}_i) \left(\bm{E} - \widetilde{\bm{u}}_i \otimes \widetilde{\bm{u}}_i \right) + \sin(q \tilde{\theta}_i) \text{skew}\left(\widetilde{\bm{u}}_i \right) + \widetilde{\bm{u}}_i \otimes \widetilde{\bm{u}}_i \right) \bm{Q}_i,
	\label{georodrigues}
\end{equation}
where $\tilde{\theta}_i$ and $\widetilde{\bm{u}}_i$ are the angle and axis of the rotation tensor $\bm{Q}_{i+1} \bm{Q}^T_i$ \cite{o2017modeling}.  To find $\tilde{\theta}_i$ and $\widetilde{\bm{u}}_i$ in terms of the model parameters $\theta_i$, $\bm{u}_i$, $\theta_{i+1}$, and $\bm{u}_{i+1}$, we use formulas (also due to Rodrigues) expressing the axis-angle parameters $\bm{u}_c, \ \theta_c$ of a composite rotation $\bm{Q}(\bm{u}_c,\theta_c) = \bm{Q}(\bm{u}_b,\theta_b)\bm{Q}(\bm{u}_a,\theta_a)$ in terms of those of the constituent rotations \cite{altmann2005rotations}:
\begin{equation}
	\cos\left(\frac{\theta_c}{2}\right) = \cos\left(\frac{\theta_b}{2}\right)\cos\left(\frac{\theta_a}{2}\right) - \sin\left(\frac{\theta_b}{2}\right)\sin\left(\frac{\theta_a}{2}\right) \left(\bm{u}_b \cdot \bm{u}_a \right), \label{rod1} 
\end{equation}
\begin{equation}
	\sin\left(\frac{\theta_c}{2}\right)\bm{u}_c = \sin\left(\frac{\theta_b}{2}\right)\cos\left(\frac{\theta_a}{2}\right)\bm{u}_b + \cos\left(\frac{\theta_b}{2}\right)\sin\left(\frac{\theta_a}{2}\right)\bm{u}_a +\sin\left(\frac{\theta_b}{2}\right)\sin\left(\frac{\theta_a}{2}\right)\left(\bm{u}_b\times \bm{u}_a\right)\label{rod2}.
\end{equation}
To apply these expressions to find $\tilde{\theta}_i$ and $\tilde{\bm{u}}_i$ we use $\bm{Q}(\bm{u},\theta)^T = \bm{Q}(\bm{u},-\theta)$ and identify $\theta_b = \theta_{i+1}$, $\bm{u}_b = \bm{u}_{i+1}$, $\theta_a = -\theta_{i}$, and $\bm{u}_a = \bm{u}_{i}$.

In the geodesic Cosserat model, the backbone curve $\bm{r}(\hat{s})$ is defined by the integral of
\begin{equation}
	\frac{\mbox{d} \bm{r}}{\mbox{d} \hat{s}} = \zeta_\alpha(\hat{s}) \bm{Q}(\hat{s}) \widehat{\bm{d}}_\alpha = \zeta_\alpha(\hat{s}) \bm{d}_\alpha(\hat{s})
\end{equation}
where summation over repeated Greek indices, indicating Cartesian components, is implied.  The formula for $\bm{r}(\hat{s})$ is
\begin{eqnarray}
	\bm{r}(\hat{s}) = \bm{r}_0 + \int_0^{\hat{s}} \zeta_\alpha(u) \bm{Q}(u)\widehat{\bm{d}}_\alpha \mbox{d}u, \label{eqrGEK}
\end{eqnarray}
where $u$ is a dummy variable of integration.  The components $\zeta_\alpha(\hat{s})$ are assumed to be constants on each segment, i.e. $\zeta_\alpha(\hat{s}) = \zeta_{i,\alpha}$ for $\hat{s}_i \leq \hat{s} < \hat{s}_{i+1}$.  The integral from $0$ to $\hat{s}$ splits up into integrals over the segments up to the one including $\hat{s}$.  If $\hat{s}$ is in segment $i$, we have for the last segment's contribution
\begin{align}
	\int_{\hat{s}_{i}}^{\hat{s}} \zeta_\alpha(u) \bm{Q}(u) \widehat{\bm{d}}_\alpha \mbox{d}u  &= \zeta_{i, \alpha} \left(\int_{0}^{q(\hat{s})}\bm{Q}_i(q(u)) \frac{\mbox{d} u}{\mbox{d} q} \mbox{d}q\right) \widehat{\bm{d}}_\alpha \nonumber \\
	&= \Bigg( \zeta_{i, \alpha} \hat{L}_{i} \int_0^{q(\hat{s})} \Big(\cos (q  \tilde{\theta}_i) \left(\bm{E} - \widetilde{\bm{u}}_i \otimes \widetilde{\bm{u}}_i \right) \nonumber \nonumber \\
	& + \sin(q  \tilde{\theta}_i) \text{skew}\left(\widetilde{\bm{u}}_i \right) + \widetilde{\bm{u}}_i \otimes \widetilde{\bm{u}}_i \Big) \mbox{d}q \Bigg)\bm{Q}_i\widehat{\bm{d}}_\alpha \nonumber \\
	&= \zeta_{i, \alpha} \hat{L}_{i} \Bigg( \Bigg(\frac{\sin( q(\hat{s}) \tilde{\theta_i})}{\tilde{\theta}_i}\Bigg) (\bm{d}_{i,\alpha} - f_{i,\alpha} \widetilde{\bm{u}}_i) \nonumber \\
	&+ \Bigg(\frac{1 -\cos( q(\hat{s}) \tilde{\theta}_i) }{\tilde{\theta}_i} \Bigg) (\widetilde{\bm{u}}_i \times \bm{d}_{i,\alpha}) + q(\hat{s}) f_{i,\alpha}\widetilde{\bm{u}}_i \Bigg), \label{GCri}
\end{align}
where $f_{i,\alpha} = \widetilde{\bm{u}}_i \cdot \bm{d}_{i,\alpha}$, and we used $\mbox{d} u/\mbox{d} q = \hat{L}_{i}$.  In the geodesic extensible Kirchoff model, $\zeta_{i,1} = \zeta_{i,2} = 0$ for all $i$, so the sum over $\alpha$ in Equation (\ref{GCri}) includes only the $\alpha = 3$ term.  Let us denote the right hand side of equation (\ref{GCri}) as $\bm{p}_i(q)$.  We then have for the final result
\begin{equation}
	\bm{r}(\hat{s}) = \bm{r}_0 + \sum_{i=0}^{i_{\hat{s}} - 1}\bm{p}_i(1) + \bm{p}_{i_{\hat{s}}}(q(\hat{s})),
	\label{GCr}
\end{equation}
where $i_{\hat{s}}$ is the index of the cylinder containing arc-length coordinate $\hat{s}$.  Thus in the geodesic Cosserat model, $\bm{r}(\hat{s})$ depends on all parameters $\theta^\text{Ax}_i, \ \beta^\text{Ax}_i$ and $\gamma^\text{Ax}_i$ for $i=0 \ldots i_{\hat{s}}+1$, $\zeta_{i,\alpha}$, for $i=0 \ldots i_{\hat{s}}$, $\alpha = 1,2,3$, as well as $\bm{r}_0$.

\subsection{Energies in the geodesic models}\label{Energies in the geodesic models}
Here we derive the filament energy $E_{i}$ in segment $i$, in which $\hat{s}_i \leq \hat{s} < \hat{s}_{i+1}$.  The total energy in the filament is a sum over the $N_\text{k}-1$ segments:
\begin{equation}
	E = \sum_{i=0}^{N_\text{k}-2} E_i,
	\label{energysum}
\end{equation}
and each $E_i$ is itself a sum over the various terms in the expression of the energy density $\varepsilon(\hat{s})$ (see Equation \ref{epsilondef} in the main text).  To evaluate $E_i$ we will need to calculate integrals of the form 
\begin{equation}
	\int_{\hat{s}_i}^{\hat{s}_{i+1}} \kappa_\alpha(\hat{s})\kappa_\beta(\hat{s}) \mbox{d}\hat{s}, \
	\int_{\hat{s}_i}^{\hat{s}_{i+1}} \sigma_\alpha(\hat{s})\sigma_\beta(\hat{s}) \mbox{d}\hat{s}, \ \text{and} \ \int_{\hat{s}_i}^{\hat{s}_{i+1}} \kappa_\alpha(\hat{s})\sigma_\beta(\hat{s}) \mbox{d}\hat{s} \nonumber
\end{equation}
using the geodesic parameterization of $\bm{r}(\hat{s})$ and $\bm{Q}(\hat{s})$.  

We start with the integrals over $\kappa_\alpha(\hat{s})\kappa_\beta(\hat{s})$.  The definition of $\kappa_\alpha(\hat{s})$ is
\begin{equation}
	\kappa_\alpha(\hat{s}) = \text{ax}\left( \bm{Q}^T(\hat{s}) \frac{\mbox{d}}{\mbox{d}\hat{s}} \bm{Q}(\hat{s})  \right) \cdot \widehat{\bm{d}}_\alpha. %ignore double superscript warning
	\label{kappaalpha}
\end{equation}
On segment $i$, we can write this in terms of the local variable $q(\hat{s};\hat{s}_i, \hat{s}_{i+1})$ as 
\begin{align}
	\kappa_{\alpha}(q(\hat{s})) &= \frac{\mbox{d} q}{\mbox{d} \hat{s}}\text{ax}\left( \bm{Q}^T_i(q(\hat{s}))\frac{\mbox{d}}{\mbox{d} q}\bm{Q}_i(q(\hat{s}))  \right) \cdot \widehat{\bm{d}}_\alpha \nonumber \\
	&= \hat{L}_i^{-1}\text{ax}\left( \bm{Q}^T_i(q)\frac{\mbox{d}}{\mbox{d} q}\bm{Q}_i(q)\right)   \cdot \widehat{\bm{d}}_\alpha.
	\label{kappaalphaq}
\end{align}
The argument of the $\text{ax}$ operation is 
\begin{align}
	\bm{Q}^T_i(q)\frac{\mbox{d}}{\mbox{d} q}\bm{Q}_i(q)  &= \bm{Q}^T_i\left(\exp\left(q \ln \widetilde{\bm{Q}}_i\right) \right)^T \frac{\mbox{d}}{\mbox{d} q}\left( \exp\left(q \ln \widetilde{\bm{Q}}_i\right) \bm{Q}_i \right) \nonumber \\
	&= \bm{Q}^T_i \left(\exp\left(q \ln \widetilde{\bm{Q}}_i\right) \right)^T \ln \widetilde{\bm{Q}}_i \exp\left(q \ln \widetilde{\bm{Q}}_i\right) \bm{Q}_i \nonumber \\
	&= \bm{Q}^T_i \ln \widetilde{\bm{Q}}_i \bm{Q}_i,
\end{align}
where $\widetilde{\bm{Q}}_i = \bm{Q}_{i+1}\bm{Q}^T_i$, and where we have used the facts that $\exp\left(q \bm{A}\right)$ and $\bm{A}$ commute for any matrix $\bm{A}$ and that $\exp\left(q \ln \widetilde{\bm{Q}}_i\right) $ is orthogonal.  A major simplification has occurred, in that the dependence on $q$ has dropped out.  Returning to equation (\ref{kappaalphaq}) we have
\begin{align}
	\kappa_{\alpha}(q) &= \hat{L}_i^{-1}\text{ax}\left( \bm{Q}^T_i \ln \widetilde{\bm{Q}}_i \bm{Q}_i \right)\cdot \widehat{\bm{d}}_\alpha \nonumber \\
	&= \hat{L}_i^{-1} \left(\bm{Q}^T_i \ \text{ax} \left(\ln \widetilde{\bm{Q}}_i \right)\right)\cdot \widehat{\bm{d}}_\alpha \nonumber \\
	&= \hat{L}_i^{-1} \text{ax} \left(\ln \widetilde{\bm{Q}}_i \right) \cdot \bm{d}_{i,\alpha},
\end{align}
where we used the identity $\text{ax}\left(\bm{Q} \bm{A} \bm{Q}^T\right) = \text{det}\left(\bm{Q}\right)\bm{Q} \ \text{ax}\left(\bm{A} \right)$ for all skew-symmetric tensors $\bm{A}$ and orthogonal tensors $\bm{Q}$, as well as the fact that the inner product is invariant under orthogonal rotations \cite{o2017modeling}.  In the axis-angle parameterization $\widetilde{\bm{Q}}_i = \exp\left(\text{skew}\left(\tilde{\theta}_i \widetilde{\bm{u}}_i\right)\right)$, and we have 
\begin{equation}
	\kappa_{\alpha}(q) = \hat{L}_i^{-1}\tilde{\theta}_i \widetilde{\bm{u}}_i \cdot \bm{d}_{i,\alpha}.
\end{equation}
Proceeding to the integral calculation, we have
\begin{align}
	\int_{\hat{s}_i}^{\hat{s}_{i+1}} \kappa_\alpha(\hat{s})\kappa_\beta(\hat{s}) \mbox{d}\hat{s} &= \int_{0}^{1} \hat{L}_i^{-2}\kappa_\alpha(q)\kappa_\beta(q) \hat{L}_i \mbox{d}q \nonumber \\
	&= \hat{L}_i^{-1}\tilde{\theta}^2_i \left(\widetilde{\bm{u}}_i \cdot \bm{d}_{i,\alpha} \right) \left(\widetilde{\bm{u}}_i \cdot \bm{d}_{i,\beta} \right).
\end{align}
Finally, we note that the vectors $\bm{d}_{i,\alpha} = \bm{Q}_i \bm{d}_\alpha$ can be expressed in terms of the model parameters, so that the final result depends only on $\mathcal{K}$ as required.

We next consider the integrals over $\sigma_\alpha(\hat{s})\sigma_\beta(\hat{s})$.  The definition of $\sigma_\alpha(\hat{s})$ is 
\begin{equation}
	\sigma_\alpha(\hat{s}) = \left(\bm{Q}^T(\hat{s}) \frac{\mbox{d}\bm{r}}{\mbox{d}\hat{s}} - \frac{\mbox{d}\widehat{\bm{r}}}{\mbox{d}\hat{s}} \right)\cdot\widehat{\bm{d}}_\alpha \nonumber
\end{equation}
We next assume that the filament has zero shear or stretch in its un-deformed configuration, so that $\mbox{d}\widehat{\bm{r}}/\mbox{d}\hat{s} = \widehat{\bm{d}}_3$, but this assumption could be relaxed.  In the geodesic parameterization of $\bm{r}(\hat{s})$, we then have
\begin{align}
	\sigma_\alpha(\hat{s}) &= \left(\bm{Q}^T(\hat{s}) \zeta_\beta(\hat{s})\bm{d}_\beta(\hat{s})\right) \cdot \widehat{\bm{d}}_\alpha - \delta_{\alpha,3} \nonumber \\
	&=  \left(\zeta_\beta(\hat{s})\bm{d}_\beta(\hat{s}) \right)\cdot \bm{d}_\alpha(\hat{s}) - \delta_{\alpha,3} \nonumber \\
	&= \zeta_\alpha(\hat{s})- \delta_{\alpha,3},
\end{align}
where $\delta_{\alpha,3}$ is the Kronecker delta, and where we have used the invariance of the inner product under orthogonal rotations as well as the orthogonality of $\bm{d}_\alpha(\hat{s})$ and $\bm{d}_\beta(\hat{s})$ for $\beta \neq \alpha$.  We take $\zeta_\alpha(\hat{s}) = \zeta_{i,\alpha}$ to be a constant on segment $i$ (although this assumption could be relaxed without overly complicating the model), so that the integrand $\sigma_\alpha(\hat{s})\sigma_\beta(\hat{s})$ becomes independent of $\hat{s}$ on the segment.  We have
\begin{align}
	\int_{\hat{s}_i}^{\hat{s}_{i+1}} \sigma_\alpha(\hat{s})\sigma_\beta(\hat{s}) \mbox{d}\hat{s} &= \int_{\hat{s}_i}^{\hat{s}_{i+1}} \left(\zeta_{i,\alpha}- \delta_{\alpha,3}\right)\left(\zeta_{i,\beta}- \delta_{\beta,3}\right) \mbox{d}\hat{s} \nonumber \\
	&= \hat{L}_i\left(\zeta_{i,\alpha}- \delta_{\alpha,3}\right)\left(\zeta_{i,\beta}- \delta_{\beta,3}\right).
\end{align}
For $\alpha = \beta = 3$, this result implies that the stretching energy is 
\begin{align}
	E_i^\text{stretch} &= \frac{S_{3,3}}{2}\int_{\hat{s}_i}^{\hat{s}_{i+1}} \sigma_3(\hat{s})^2 \mbox{d}\hat{s} \nonumber \\
	&= \frac{S_{3,3}}{2} \hat{L}_i\left(\zeta_{i,3} - 1\right)^2 \nonumber \\
	&= \frac{S_{3,3}}{2\hat{L}_i}\left(L_i - \hat{L}_i\right)^2, \label{GCstretchenergy}
\end{align}
where we used the fact that $\zeta_{i,3} = L_i / \hat{L}_i$.  Thus the stretching energy is that of a harmonic spring with a spring constant given by $S_{3,3}/\hat{L}_i =E_\text{mod} A/\hat{L}_i $ (see the Parameterization section of the Supplementary Material).  We also have that the shearing energy is
\begin{equation}
	E_i^\text{shear} = \frac{S_{1,1}}{2} \hat{L}_i \left(\zeta_{i,1}^2 + \zeta_{i,2}^2 \right).
\end{equation}

We note that stretching energy, Equation \ref{GCstretchenergy}, is equivalent to the stretching energy used in the MEDYAN model, Equation \ref{MEDstretch} in the main text.  As a matter of interest, we next show that the MEDYAN bending energy, Equation \ref{MEDbend} in the main text, agrees to second order in $\theta_{i,i+1}^\text{MED}$ with the bending energies in the geodesic models if there is no filament twist, in which case $\widetilde{\bm{u}}_i \cdot \bm{d}_{i,3} = 0$. The bending energy in the geodesic model becomes
\begin{align}
	E_i^\text{bend} &= \frac{B_{1,1}}{2} \int_{\hat{s}_i}^{\hat{s}_{i+1}} (\kappa_1(\hat{s})^2 + \kappa_2(\hat{s})^2) \mbox{d}\hat{s} \nonumber \\
	&= \frac{B_{1,1}}{2 \hat{L}_i} \tilde{\theta}_i^2 \left( (\widetilde{\bm{u}}_i \cdot \bm{d}_{i,1})^2 + (\widetilde{\bm{u}}_i \cdot \bm{d}_{i,2})^2 \right) \nonumber \\
	&= \frac{B_{1,1}}{2 \hat{L}_i} \tilde{\theta}_i^2,
	\label{eq:si-geodesic-bend}
\end{align}
where the third line follows since $\lvert\lvert\widetilde{\bm{u}}_i \rvert \rvert =1$.  If we let $\theta_{i,i+1}^\text{MED} = \tilde{\theta}_i$, and expand Equation \ref{MEDbend} to second other with respect to $\theta_{i,i+1}^\text{MED}$, then Equation \ref{MEDbend} in the main text and Equation \ref{eq:si-geodesic-bend} are equivalent.

Finally, through similar steps to those outlined above it can be shown that the integral over $\kappa_\alpha(\hat{s})\sigma_\beta(\hat{s})$ is
\begin{equation}
	\int_{\hat{s}_i}^{\hat{s}_{i+1}} \kappa_\alpha(\hat{s})\sigma_\beta(\hat{s}) \mbox{d}\hat{s} = \tilde{\theta}_i \left(\widetilde{\bm{u}}_i \cdot \bm{d}_{i,\alpha}\right)\left(\zeta_{i,\beta}- \delta_{\beta,3}\right).
\end{equation}
This completes the derivation of the filament energy, expressed in terms of the free model parameters in $\mathcal{K}$.

\subsection{MEDYAN Implementation}

To implement the GC model into a network-level simulation platform such as MEDYAN, several additional modelling choices need to be considered, particularly related to how chemical reactions such as (de)polymerization and binding of cross-linkers and molecular motors will occur.  Other steps for implementation are necessary, such as finding explicit expressions for the energy gradient functions to use in our custom numerical minimization routine, but we omit here these tedious but straightforward details.  In the remainder of this section we describe how chemical reactions are handled in our implementation, but we first give a brief overview of the MEDYAN simulation platform.

\subsubsection{MEDYAN simulation protocol}

A detailed introduction to the MEDYAN (Mechanochemical Dynamics of Active Networks) model can be found in Ref. \citenum{popov2016medyan}, and several applications can be found in Refs. \citenum{floyd2019quantifying, chandrasekaran2019remarkable, komianos2018stochastic, ni2019turnover, li2020tensile, floyd2020gibbs,ni2021membrane,floyd2021segmental,floyd2021understanding}.  Here we describe the aspects of MEDYAN relevant to the this paper, and direct the reader to the above references for a thorough description.  A MEDYAN simulation proceeds by iterating a cycle of four steps which propagate the chemical and mechanical dynamics forward while coupling between the two.  The steps are as follows:
\begin{enumerate}
	\item Evolve system using stochastic chemical simulation for a time $\delta t$.
	\item Compute the changes in the mechanical energy resulting from the reactions that occurred in step 1).
	\item Mechanically equilibrate the network in response to the new stresses from step 2).  
	\item Update the reaction rates of force-sensitive reactions based on the new tensions from step 3).  
\end{enumerate}
The mechanics of the system consists of a filament mechanical model, which is the primary subject of the this paper, as well as other potentials describing the stretching of cross-linkers and motors and the excluded volume repulsion between nearby filaments and between filaments and the boundary.  These latter potentials are treated here identically to previous MEDYAN works, and we refer the reader to Ref. \citenum{popov2016medyan} for a description.  We focus next on the chemical simulation protocol, step 1) of the above simulation cycle.

\subsubsection{Chemical dynamics in MEDYAN}

In MEDYAN, diffusing chemical species have discrete copy numbers and belong to several compartments that form a regular grid comprising the simulation volume.  The compartment size is chosen so that the well-mixed assumption holds inside each compartment, allowing the use of mass-action kinetics to determine propensities for participating in chemical reactions within compartments and diffusion events between adjacent compartments.  The Next Reaction Method (NRM) is used to stochastically choose which event will occur next and the time to that reaction \cite{bernstein2005simulating, gillespie1977exact}.  The user specifies the chemical species and the reactions in which they participate.  Several types of reactions are possible.  Polymerization reactions cause the subtraction of a diffusing monomer from the local compartment and its conversion into a filament species, lengthening the filament, and depolymerization reactions do the opposite.  Filaments in MEDYAN have explicit spatial coordinates rather than just the compartment-level copy numbers of the diffusing species.  This network of spatially resolved filaments lies over the compartment grid, so that sections of filaments are able to react with diffusing species according to the local compartment copy numbers.  As a result, a filament may react with a diffusing species such as a cross-linker (e.g. $\alpha$-actinin), branching (e.g. Arp2/3), or molecular motor (e.g. NMIIA) which will in turn alter the system's mechanical energy.  Binding reactions occur on a discrete set of binding sites along the filament and stochastically occur according to the number of those binding sites and the local copy number of diffusing binding molecules.  A bound molecular motor may undergo a walking reaction in which it moves one of its ends to an adjacent filament binding site, stretching the motor and generating forces.  Unbinding and motor walking reactions are modeled as force-sensitive, such that their propensities depends on the forces sustained by the molecules.  Other reactions not used in this paper but allowed in MEDYAN include filament nucleation, filament destruction, filament severing, and filament branching reactions. 

\subsubsection{Binding of linkers and motors in the GC model}

In the original MEDYAN implementation where filaments are 1D objects, the binding sites to which cross-linkers and molecular motors attach on the filaments are a discrete set of points on the 1D filament backbones.  Binding reactions are allowed when a pair of such binding sites on nearby filaments are within a user-specified distance threshold determined by the binding molecules size, and the reaction then occurs stochastically through the NRM algorithm.  In the finite-width filament models presented in this paper, binding sites are not restricted to lie on the filament backbone but instead can lie on the filament surface, which introduces an additional degree of freedom $\phi^\text{b}$ at the filament backbone position $\mathbf{r}^\text{b}$ (see Figure~\ref{CrossSection} of the main text).  We next describe two ways to determine the new degree of freedom $\phi^\text{b}$, though others may be designed as well.

The first method one can use to determine $\phi_A^\text{b}$ and $\phi_B^\text{b}$, the binding angles on the filaments $A$ and $B$ which are participating in the binding reaction, is to choose them so that they minimize the distance between the two binding sites.  This amounts to choosing the closest distance between the perimeters of two circles which are arbitrarily oriented in 3D space.  The benefit of this choice is that it ensures that when the binding happens the binding molecule does not erroneously pass through the either of the filaments, which should be sterically prohibited.  The downside is that it does not encode any microstructural information which may be useful to realistically model a helical filament like actin.  The second method which can be used to determine $\phi_A^\text{b}$ and $\phi_B^\text{b}$ is to require that all binding sites lie on one or several fixed helices wrapping around the filament.  For instance, if the binding of some molecule is known to occur on the major groove of an actin filament, then it is of interest to ensure that the angles $\phi^\text{b}$ are chosen to correspond to the location of this groove at the point $\mathbf{r}^\text{b}$.  This method could allow for sterically prohibited overlap between the binding molecule and the filament at the time of binding, but it has the benefit of encoding microstructural detail into the model.  Sterically prohibited overlap can be discouraged during the energy minimization step by including the energy penalty term $E_\text{steric}$ defined in Equation \ref{stericU}.

\subsubsection{(De)polymerization in the GC model}

Here we consider how to update the parameters describing the geodesic filament configuration when a polymerization or depolymerization reaction occurs.  We will generally have some set of model parameters before the event $\mathcal{K}^b$ and a set of parameters after the event $\mathcal{K}^a$, and the goal is to find $\mathcal{K}^a$ as a function of $\mathcal{K}^b$ depending on the type of event that occurs.  We specify the condition for determining $\mathcal{K}^a$ by requiring that the new curve does not differ on the original domain from the previous curve, so that the new curve just extends the domain of the previous one.  We fix the maximum length which a filament segment can have at $\hat{L}_j^\text{max}$, which complicates the situation by requiring slightly different update rules depending on whether the (de)polymerization event causes a change in length that passes through this maximum length.  Additionally, we need to consider separately reactions occurring at the plus and minus ends of the filament.  We will describe these various cases in turn.

The first case is of a polymerization event at the filament plus-end on a segment that is not yet at its maximum length $\hat{L}_j^\text{max}$, where we denote the segment index where the polymerization event occurs $j$.  We define the GC parameter sets $\mathcal{M}_j = \{ \theta^\text{Ax}_j, \beta^\text{Ax}_j, \gamma^\text{Ax}_j \}$ to represent the angles fixing the rotation matrix $\mathbf{Q}_j$, and $\mathcal{L}_j = \{\zeta_{j,1}, \zeta_{j,2}, \zeta_{j,3} \}$ to represent the expansion coefficients the segment $j$.  The original total filament length is $\hat{L}^b$ and that of the segment is $\hat{L}^b_j$, and after the event the lengths are respectively $\hat{L}^a = \hat{L}^b + \delta l$ and $\hat{L}_j^a = \hat{L}_j^b + \delta l$.  None of the parameters on the segments previous to $j$ will be altered.  The only parameters which will change due to this event are in $\mathcal{L}_j$ and $\mathcal{M}_{j+1}$ (which specifies the rotation matrix at  the end of segments $j$).  We will assume that $\mathcal{L}^a_j = \mathcal{L}^b_j$, i.e. that the shearing and stretching strain on the segment does not change due to the polymerization.  The choice of $\mathcal{M}^a_{j+1}$ will be made based on the condition that the tangent vector to the backbone at the previous plus-end point does not change due to the polymerization event: 
\begin{equation}
	\partial_{\hat{s}} \mathbf{r}(\hat{L}^b ; \mathcal{K}^b) = \partial_{\hat{s}} \mathbf{r}(\hat{L}^b ; \mathcal{K}^a).
\end{equation}
The choice of evaluating this condition at the previous plus-end is arbitrary, and it could be done anywhere on the segment.  This condition implies that
\begin{equation}
	\zeta_{j,\alpha}^b \mathbf{Q}_j^{b}(\hat{L}^b) \widehat{\mathbf{d}}_\alpha = \zeta_{j,\alpha}^a \mathbf{Q}_j^{a}(\hat{L}^b) \widehat{\mathbf{d}}_\alpha
\end{equation}
The original $q$ coordinate at the plus-end is $q^b = 1$, and afterwards the coordinate for $\hat{L}_j^b$ is $q^a = \frac{\hat{L}_j^b}{\hat{L}_j^b+\delta l}$.  Because the $\zeta_\alpha$ parameters are assumed equal, the above equation simplifies to
\begin{equation}
	e^{q^b \text{skew}(\tilde{\theta}_j^b \widetilde{\mathbf{u}}_j^b)}\mathbf{Q}_j = e^{q^a \text{skew}(\tilde{\theta}_j^a \widetilde{\mathbf{u}}_j^a)}\mathbf{Q}_j,
\end{equation}
or 
\begin{equation}
	\frac{q^b}{q^a}\text{skew}(\tilde{\theta}_j^b \widetilde{\mathbf{u}}_j^b) = \text{skew}(\tilde{\theta}_j^a \widetilde{\mathbf{u}}_j^a).
\end{equation}
This simplifies to 
\begin{equation}
	\frac{q^b}{q^a}\tilde{\theta}_j^b \widetilde{\mathbf{u}}_j^b = \tilde{\theta}_j^a \widetilde{\mathbf{u}}_j^a,
\end{equation}
and, since $\widetilde{\mathbf{u}}_j^b$ and $\widetilde{\mathbf{u}}_j^a$ are both unit vectors they must be equal if they point in the same direction, leaving us with
\begin{equation}
	\widetilde{\mathbf{u}}_j^b = \widetilde{\mathbf{u}}_j^a
\end{equation}
and 
\begin{equation}
	\tilde{\theta}_j^a = \tilde{\theta}_j^b\left(1 + \frac{\delta l}{\hat{L}_j^b} \right).
	\label{tildethetaapeincomplete}
\end{equation}
These equations needs to be solved to give $\theta_{j+1}^a$ in terms of the parameters in $\mathcal{M}^b_{j}$ and $\mathcal{M}^b_{j+1}$.  This can be done by first writing
\begin{equation}
	\mathbf{Q}_{j+1}^{a} = e^{\text{skew}( \tilde{\theta}_j^a\widetilde{\mathbf{u}}_j^a)}\mathbf{Q}_{j}^{b} = e^{\text{skew}( \frac{q^b}{q^a}\tilde{\theta}_j^b\widetilde{\mathbf{u}}_j^b)}\mathbf{Q}_{j}^{b}.
\end{equation}
Since the right hand side is in terms of the previous, known model parameters, it can be evaluated and a routine for then determining $\theta_{j+1}^a, \ \beta_{j+1}^a, \ \gamma_{j+1}^a$ from the resulting tensor elements can be used.  If instead of a polymerization event at the plus- end incomplete segment there were a depolymerization event, the same results would carry through except the ratio $\frac{q^b}{q^a}$ is now $\left(1 - \frac{\delta l}{\hat{L}_j^b} \right)$.

We next consider a polymerization event at the plus end which produces a new segment, when $\hat{L}_j^b = \hat{L}_j^\text{max}$.  The same condition, that the tangent at the previous plus-end should be unchanged, can be used in this case, but here it actually does not uniquely specify what the new parameters $\mathcal{M}^a_{j+2}$ should be.  The tangent can be written as
\begin{equation}
	\partial_{\hat{s}} \mathbf{r}(\hat{L}^b ; \mathcal{K}^b) = \zeta_{j,\alpha}^b \mathbf{Q}_{j+1}^{b} \widehat{\mathbf{d}}_\alpha,
\end{equation}
which depends only on $\mathcal{M}_{j+1}^b$ and $\mathcal{L}_j$.  A second order derivative on $\mathbf{r}(\hat{L}^b ; \mathcal{K}^b)$ could be use to constrain $\mathcal{M}^a_{j+2}$, but we can take the simpler option of simply setting $\mathcal{M}^a_{j+2} = \mathcal{M}^b_{j+1} = \mathcal{M}^a_{j+1}$, so that the new segment has the same rotation matrix as the previous plus-end point.  We can also take $\mathcal{L}_{j+1}^a = (0,0,1)$.  Thus the new segment is assumed to continue straight in an un-strained way from the tangent at $\hat{L}^b$.  Smoothness is still guaranteed at $\hat{L}^b$.  This freedom of parameter choice when creating a new segment is a qualitative difference compared to polymerization events occurring on an incomplete segment, but it should not introduce any serious issues into the simulation.  Any unrealistic choices for this polymerization process will be resolved during the subsequent energy minimization routine, and the effects of this on the system dynamics will be minor.  Depolymerization events causing the destruction of a plus-end segment can be trivially handled by keeping all parameters the same and simply deleting the ones for the depolymerized segment.

We next consider a polymerization event occurring at an incomplete minus-end segment.  This requires updating $\mathcal{M}_0, \ \mathcal{L}_0$, and $\mathbf{r}_0$.  We again take $\mathcal{L}_0^a = \mathcal{L}_0^b$ for simplicity.  We require the tangent vector to the backbone at the previous minus-end ($q^b = 0$) to be equal to the tangent in the new segment at $q^a = \frac{\delta l}{\hat{L}^b_0 + \delta l}$.  This amounts to the condition
\begin{equation}
	\zeta_{0,\alpha}^b \mathbf{Q}_0^{b}\widehat{\mathbf{d}}_\alpha = \zeta_{0,\alpha}^a\mathbf{Q}_0^{a}(q^a)\widehat{\mathbf{d}}_\alpha
\end{equation}
or 
\begin{equation}
	\mathbf{Q}_0^{b} = \mathbf{Q}_0^{a}(q^a) = e^{q^a \text{skew}(\tilde{\theta}_0^a\widetilde{\mathbf{u}}_0^a)}\mathbf{Q}_0^{a}.
\end{equation}
It will be easier to write the original tensor curve $\mathbf{Q}_0^{b}(q^b)$ in the opposite direction, as the curve that goes from $\mathbf{Q}_{1}^{ b}$ to $\mathbf{Q}_{0}^{b}$, written in terms of the original $q^b$ as 
\begin{equation}
	\mathbf{Q}_0^{b}(q^b) = e^{q^b \text{skew}(\tilde{\theta}_0^b \widetilde{\mathbf{u}}_0^b)}\mathbf{Q}_0^{ b}= e^{(1-q^b)\text{skew}(-\tilde{\theta}_0^b \widetilde{\mathbf{u}}_0^b)}\mathbf{Q}_1^{ b}.
\end{equation}
The tangent condition at the original minus-end $q^b = 0$ then reads
\begin{equation}
	e^{\text{skew}(-\tilde{\theta}_0^b \widetilde{\mathbf{u}}_0^b)}\mathbf{Q}_1^{ b} = e^{(1-q^a)\text{skew}(-\tilde{\theta}_0^a \widetilde{\mathbf{u}}_0^a)}\mathbf{Q}_1^{b}
\end{equation}
or 
\begin{equation}
	e^{\text{skew}(-\tilde{\theta}_0^b \widetilde{\mathbf{u}}_0^b)}=e^{(1-q^a)\text{skew}(-\tilde{\theta}_0^a \widetilde{\mathbf{u}}_0^a)}.
\end{equation}
This leads to 
\begin{equation}
	\widetilde{\mathbf{u}}_0^b = \widetilde{\mathbf{u}}_0^a
\end{equation}
and 
\begin{equation}
	\tilde{\theta}_0^b = (1-q^a)\tilde{\theta}_0^a,
\end{equation}
or 
\begin{equation}
	\tilde{\theta}_0^a = \frac{\hat{L}_0^b +\delta l}{\hat{L}_0^b}\tilde{\theta}_0^b.
\end{equation}
Writing
\begin{equation}
	\mathbf{Q}_0^{ a} = e^{\text{skew}(-\tilde{\theta}_0^a \widetilde{\mathbf{u}}^a_0)} \mathbf{Q}_1^{ b}
\end{equation}
and substituting for $\tilde{\theta}_0^a$ and $\widetilde{\mathbf{u}}^a_0$ in terms of their original counterparts $\tilde{\theta}_0^b$ and $\widetilde{\mathbf{u}}^b_0$, this expression can be evaluated, and the sought after model parameters $\theta_0^a, \ \beta_0^a, \ \gamma_0^a$ can be found from the tensor $\mathbf{Q}_0^{ a}$.  To find $\mathbf{r}_0^a$, the new minus-end position of the backbone, we require that using the new parameters $\mathcal{K}^a$ the position of the backbone at $\hat{s} = \delta l$ is equal to the previous minus-end backbone position, i.e. that
\begin{equation}
	\mathbf{r}_0^b = \mathbf{r}^a(\delta l) = \mathbf{r}_0^a + \int_0^{\delta l} \partial_{\hat{s}} \mathbf{r}(\hat{s} ; \mathcal{K}^a) d\hat{s}
\end{equation}
or 
\begin{equation}
	\mathbf{r}_0^a = \mathbf{r}_0^b - \mathbf{p}_0\left(\frac{\delta l}{\hat{L}_0^b + \delta l}; \mathcal{L}_0^a, \mathcal{M}_0^a, \mathcal{M}_1^a\right),
	\label{newminusendpoly}
\end{equation}
where $\mathbf{p}_0(q)$ is defined in Equation \ref{GCr}.

For a depolymerization event at an incomplete minus-end segment, we evaluate the tangent condition at $q^a = 0$, $q^b = \frac{\delta l}{\hat{L}_0^b}$.  Through similar steps to those outlined above, this can be shown to give the condition
\begin{equation}
	(1-q^b) \tilde{\theta}_0^b =  \tilde{\theta}^a_0.
\end{equation}
The new minus-end position can be found as the previous backbone position evaluated at $\hat{s} = \delta l$:
\begin{equation}
	\mathbf{r}_0^a = \mathbf{r}(\delta l; \mathcal{L}_0^b, \mathcal{M}_0^b, \mathcal{M}_1^b).
	\label{newminusenddepoly}
\end{equation}

If the polymerization event at the minus-end creates a new segment, we follow the same steps as in the plus-end new segment case by allowing the new segment to continue the tangent at the previous minus-end in an unstrained way.  This amounts to $\mathcal{M}_0^a = \mathcal{M}_0^b = \mathcal{M}_1^a$ and $\mathcal{L}_0 = (0,0,1)$.  The new minus-end position $\mathbf{r}_0^a$ is found using Equation \ref{newminusendpoly}.  Finally if a depolymerization event destroys a segment, we simply find the new minus-end position using Equation \ref{newminusenddepoly} and then discard the parameters of the destroyed segment, keeping all other parameters the same.

\vskip 5mm

\leftline{{\bf Acknowledgements:}}
\noindent
This work was supported by the Engineering and Physical Sciences Research Council, grant number EP/V047469/1, awarded to Radek Erban.  This work was also supported by the National Science Foundation, grant number CHE-210268 and a Visiting Research Fellowship from Merton College, Oxford, awarded to Garegin Papoian.

%\bibliography{MP.bib}
\bibliographystyle{unsrt}

\providecommand{\latin}[1]{#1}
\makeatletter
\providecommand{\doi}
{\begingroup\let\do\@makeother\dospecials
	\catcode`\{=1 \catcode`\}=2 \doi@aux}
\providecommand{\doi@aux}[1]{\endgroup\texttt{#1}}
\makeatother
\providecommand*\mcitethebibliography{\thebibliography}
\csname @ifundefined\endcsname{endmcitethebibliography}
{\let\endmcitethebibliography\endthebibliography}{}

%\end{thebibliography}

\end{document}